\documentclass[twocolumn,resetfootnote]{aastex701}
\usepackage[T1]{fontenc}
\usepackage[utf8]{inputenc}
\let\tablenum\relax
\usepackage{mathtools}
\usepackage{siunitx}
\usepackage{physics}
\usepackage{graphicx}
\usepackage{tabularray}
\usepackage{colortbl}
\usepackage{enumerate}
\usepackage{url}
\usepackage{comment}

\newcommand{\Msun}{\mathrm{M_\odot}}
\newcommand{\Zsun}{\mathrm{Z_\odot}}
\DeclareSIUnit\angstrom{\text {Å}}

\begin{document}

\title[Pop~III stars where you least expect them]{Catching the Nebular Needle in a Polluted Haystack:
Line-emission Signatures from Population~III-forming Pockets around Massive Galaxies at the End of Reionization}

\correspondingauthor{Alessandra Venditti}
\author[0000-0003-2237-0777]{Alessandra Venditti}
\altaffiliation{Cosmic Frontier Center Prize Fellow}
\email{alessandra.venditti@utexas.edu}
\affiliation{Department of Astronomy, University of Texas at Austin, 2515 Speedway, Stop C1400, Austin, TX 78712, USA}
\affiliation{Cosmic Frontier Center, The University of Texas at Austin, Austin, TX 78712}

\author[0000-0002-9231-1505]{Luca Graziani}
\affiliation{Dipartimento di Fisica, Sapienza, Università di Roma, Piazzale Aldo Moro 5, 00185, Roma, Italy}
\affiliation{INFN, Sezione di Roma I, Piazzale Aldo Moro 2, 00185, Roma, Italy}
\affiliation{INAF-Osservatorio Astronomico di Roma, Via di Frascati 33, 00078, Monte Porzio Catone, Italy}
\email{luca.graziani@uniroma1.it}

\author[0000-0001-9317-2888]{Raffaella Schneider}
\affiliation{Dipartimento di Fisica, Sapienza, Università di Roma, Piazzale Aldo Moro 5, 00185, Roma, Italy}
\affiliation{INFN, Sezione di Roma I, Piazzale Aldo Moro 2, 00185, Roma, Italy}
\affiliation{INAF-Osservatorio Astronomico di Roma, Via di Frascati 33, 00078, Monte Porzio Catone, Italy}
\email{raffaella.schneider@uniroma1.it}

\author[0000-0003-0212-2979]{Volker Bromm}
\affiliation{Department of Astronomy, University of Texas at Austin, 2515 Speedway, Stop C1400, Austin, TX 78712, USA}
\affiliation{Weinberg Institute for Theoretical Physics, University of Texas at Austin, Austin, TX 78712, USA}
\email{vbromm@astro.as.utexas.edu}

\author[0000-0002-8984-0465]{Julian B. Muñoz}
\affiliation{Department of Astronomy, University of Texas at Austin, 2515 Speedway, Stop C1400, Austin, TX 78712, USA}
\email{julianbmunoz@utexas.edu}

\author[0000-0003-1408-7373]{Claudia Di Cesare}
\email{claudia.dicesare@ista.ac.at}
\affiliation{Institute of Science and Technology Austria, Am Campus 1, 3400 Klosterneuburg, Austria}

\author[0000-0003-3050-1765]{Rosa Valiante}
\email{rosa.valiante@inaf.it}
\affiliation{INAF-Osservatorio Astronomico di Roma, Via di Frascati 33, 00078, Monte Porzio Catone, Italy}

\author[0000-0003-2536-1614]{Antonello Calabrò}
\email{antonello.calabro@inaf.it}
\affiliation{INAF-Osservatorio Astronomico di Roma, Via di Frascati 33, 00078, Monte Porzio Catone, Italy}

\author[0000-0002-4985-3819]{Roberto Maiolino}
\email{rm665@cam.ac.uk}
\affiliation{Kavli Institute for Cosmology, University of Cambridge, Madingley Road, Cambridge CB3 0HA, UK}
\affiliation{Cavendish Laboratory, University of Cambridge, 19 JJ Thomson
Avenue, Cambridge CB3 0HE, UK}
\affiliation{Department of Physics and Astronomy, University College London,
Gower Street, London WC1E 6BT, UK}

\author[0000-0001-8519-1130]{Steven L. Finkelstein}
\email{stevenf@astro.as.utexas.edu}
\affiliation{Department of Astronomy, University of Texas at Austin, 2515 Speedway, Stop C1400, Austin, TX 78712, USA}
\affiliation{Cosmic Frontier Center, The University of Texas at Austin, Austin, TX 78712}

\author[0000-0002-9729-3721]{Massimiliano Parente}
\email{parente.m@ufl.edu}
\affiliation{Department of Astronomy, University of Florida, 211 Bryant Space Sciences Center, Gainesville, FL 32611, USA}

\author{Matteo Saggini}
\email{matteo.saggini@uniroma1.it}
\affiliation{Dipartimento di Fisica, Sapienza, Università di Roma, Piazzale Aldo Moro 5, 00185, Roma, Italy}
\affiliation{Dipartimento di Fisica, Tor Vergata, Università di Roma, Via Cracovia 50, 00133, Roma, Italy}
\affiliation{INFN, Sezione di Roma I, Piazzale Aldo Moro 2, 00185, Roma, Italy}

\author[0000-0002-0302-2577]{John Chisholm}
\email{chisholm@austin.utexas.edu}
\affiliation{Department of Astronomy, University of Texas at Austin, 2515 Speedway, Stop C1400, Austin, TX 78712, USA}
\affiliation{Cosmic Frontier Center, The University of Texas at Austin, Austin, TX 78712}

\begin{abstract}
Finding the first generation of (Population III or Pop III) stars is one of the most ambitious and exciting challenges of astrophysics. JWST opened concrete prospects for their detection during the Epoch of Reionization (EoR), where increasing evidence suggests that residual Pop III formation may persist, even within pristine pockets of high-mass halos, due to inhomogeneous enrichment. However, the identification of Pop III stars within globally enriched environments will be challenging. We investigate the detectability of a subdominant Pop III component in/around massive ($M_\star \gtrsim 10^9 ~\Msun$) galaxies at $z \approx 6.5 - 9$ from the \texttt{dustyGadget} cosmological simulation suite, and the confusion arising from second-generation (Pop II) stars in their surroundings. We find that young ($\lesssim 1$ Myr), massive ($M_\mathrm{III} \sim 6 \times 10^5 ~\Msun$) Pop III clusters forming within these galaxy environments are responsible for strong HeII1640 line emission ($L_\mathrm{HeII1640} \gtrsim 10^{41} ~\si{erg.s^{-1}}$), which would be detectable with $\approx 10 (50)$ h of medium-resolution observations with NIRSpec/IFU at $z \approx 6 (10)$. These bright luminosities cannot be produced by standard Pop II populations alone. On the other hand, the dominant Pop II component within massive ``hybrid'' Pop III hosts powers strong metal line emission ($L_\mathrm{[OIII]5007} \gtrsim 10^{42} ~\si{erg.s^{-1}}$), indicating that the detection of metal lines alone cannot exclude the presence of Pop IIIs in high-$z$ galaxy environments. We further discuss candidate selection strategies based on Ly$\alpha$, H$\alpha$ and H$\beta$ emission, and how spatially resolved observations may enable the detection of isolated, pristine pockets in the outskirts of massive halos.
\end{abstract}

\keywords{
\uat{Population III stars}{1285} --- \uat{Reionization}{1383} --- \uat{High-redshift galaxies}{734} --- \uat{Early universe}{435} --- \uat{James Webb Space Telescope}{2291} --- \uat{Theoretical models}{2107} --- \uat{Hydrodynamical simulations}{767} --- \uat{Photoionization}{2060} --- \uat{H II regions}{694} --- \uat{Emission line galaxies}{459}
}

\section{Introduction}
\label{sec:introduction}

Population ~III (Pop~III) stars constitute the first stellar generation formed from metal-free gas, which reflects the composition of the primordial Universe produced by the Big Bang Nucleosynthesis \citep{Alpher_1948, Burbidge_1957}. The overall consensus is that the bulk of Pop~III formation occurs in pristine minihalos ($\sim 10^5 - 10^6 ~\Msun$) at Cosmic Dawn ($z \sim 20 - 30$, \citealt{Bromm_2013, Klessen_Glover_2023}). In these environments, the collapse of gas clouds is fueled by inefficient H$_2$ cooling, resulting in significantly lower star-formation efficiencies (SFEs) and more top-heavy initial mass functions (IMFs) with respect to present-day stars \citep[e.g.,][]{Hirano_2014}. Once formed, these massive Pop~III stars synthesize metals and release them into the surrounding medium after a short lifetime, so that subsequent generations of Pop~II stars can eventually form from the enriched gas. 

However, a host of numerical simulations \citep{Tornatore_2007_PopIII, Maio_2010, Johnson_2013, Pallottini_2014, Xu_2016_latePopIII, Xu_2016_XRB, Jaacks_2019, Sarmento_2018, Sarmento_Scannapieco_2022, Sarmento_Scannapieco_2025, Skinner_Wise_2020, Zier_2025} and semi-analytical models \citep{Visbal_2020, Munoz_2021, Cruz_2025, Ventura_2024} suggest that this simple picture of an early Pop~III phase quickly transitioning to Pop~II formation may be too simplistic. In fact, metal enrichment is a highly inhomogeneous process, that can leave patches of pristine gas capable of forming Pop~III stars at later times, possibly down to the EoR and beyond. Some models even predict that Pop~III stars at late times could form in halos much larger than the first minihalos \citep{Liu_Bromm_2020, Bennet_Sijacki_2020, Riaz_2022, Storck_2025}, including globally enriched halos \citep{Venditti_2023, Hegde_Furlanetto_2025}, and studies suggest that externally irradiated halos within the atomic-cooling regime may host more efficient star formation due to the onset of efficient HD cooling (\citealt{Greif_Bromm_2006, Greif_2008, Bromm_2009}, and, more recently \citealt{Sugimura_2024, Jeong_2026}).

Finding faint star-formation sites at Cosmic Dawn is extremely challenging \citep[e.g.,][]{Schauer_2020}. JWST has now opened an unprecedented observational window into the first billion year of cosmic history, with the earliest spectroscopically confirmed galaxies to date reaching as far as $z \approx 14.44$ \citep{Naidu_2025}. And yet, only one candidate Pop~III-dominated system has been reported so far at $z > 10$ \citep{Maiolino_2024, Maiolino_2026, Ubler_2026}, while a potential top-heavy IMF component -- typical of Pop~IIIs -- has been proposed to alleviate the tension between pre-JWST galaxy formation models and the observed abundance of UV-bright galaxies at these early times \citep{Inayoshi_2022, Finkelstein_2023, Harikane_2023, Harikane_2024, Yung_2024, Trinca_2024, Ventura_2024, Cueto_2024, Hutter_2025, Lu_2025, Harvey_2025, Jeong_2025, Mauerhofer_2025}\footnote{\citet{Jeon_2026} recently suggested an alternative and promising strategy to reveal early star-forming minihalos, by capturing the bright signal of pair-instability supernovae (PISNe) in overdense fields with an accelerated star-formation history.}. On the other hand, candidate Pop~III systems have been discovered at later cosmic epochs, including the EoR \citep{Vanzella_2020, Vanzella_2023, Nakajima_2025, Wang_2024, Cullen_2025, Morishita_2025}, and even the post-reionization era \citep{Cai_2025, Mondal_2025}; an early analysis by \citet{Jimenez_Haiman_2006} further suggested that a fraction of $\sim 20 - 30\%$ of Pop~III stars could explain the HeII1640 and Ly$\alpha$ emission observed in a stacked galaxy sample at $z \sim 3$ \citep{Steidel_2001}.
Finally, the study of \citet{Venditti_2025} indicates that efficient Pop~III formation (possibly within heavy, atomic-cooling halos) would be required to explain bright Pop~III candidates such as the AMORE6 galaxy at $z \approx 5.7$, based on preliminary constraints on the Pop~III UV luminosity function at $z \sim 6$ \citep{Fujimoto_2025, Fujimoto_2025_revisited}.

Most of these candidates have been identified by relying on a combination of low-metallicity and spectral-hardness diagnostics \citep{Inoue_2011, Zackrisson_2011, Mas-Ribas_2016, Nakajima_Maiolino_2022, Katz_2023, Trussler_2023, Cleri_2023}. 
However, the recent study of \citet{Rusta_2025} suggests that strong metal-lines may arise even from Pop~III-dominated galaxies, due to efficient self-enrichment of the surrounding gas on very short timescales, after the most massive stars explode as supernovae (SNe)\footnote{A similar effect may occur in high-$z$ IGM absorption signatures around Pop~III hosting halos \citep{Wang_GRB2012}.}. Establishing a complete absence of metals presents a significant observational challenge, due to the limited sensitivity of our instruments \citep[e.g.,][]{Frebel_2009}. In practice, even the lowest available metallicity constraint at high redshifts ($< 1.2 \times 10^{-3} ~\Zsun$, derived for the AMORE6 galaxy based on the non-detection of metal lines, \citealt{Morishita_2025}) still lies at least one order of magnitude above the most metal-poor star observed in the local Universe (i.e., the red giant star SDSS J0715-7334 identified in the Large Magellanic Cloud, with a measured metallicity $\sim (4.2 - 11) \times 10^{-5} ~\Zsun$, \citealt{Ji_2025}).

On the other hand, massive, metal-poor stars should produce intense He$^+$-ionizing radiation, leaving a distinctive mark in the spectra of high-$z$ galaxies. Early works \citep{Tumlinson_Shull_2000, Tumlinson_2001, Bromm_2001_spectra, Oh_2001} first pointed out the exceptionally strong HeII lines (e.g., HeII1640 and HeII4686) powered by the cascade recombination of HeIII in the medium surrounding massive ($\gtrsim 40 ~\si{M_\odot}$) Pop~III stars. Later models for massive Pop~III stars and stellar populations with zero \citep{Schaerer_2002} or extremely low metallicity \citep{Schaerer_2003, Raiter_2010} confirmed this prediction, albeit with more conservative estimates of the resulting HeII line luminosities.
Strong Ly$\alpha$ lines and Balmer recombination lines are also expected \citep[e.g.][]{Inoue_2011, Mas-Ribas_2016, Nakajima_Maiolino_2022}, as well as steep Balmer jumps and UV slopes, due to powering of the nebular-continuum portion of the spectrum from hot, ionized gas in the vicinity of the stars \citep{Zackrisson_2011, Trussler_2023}.

While these diagnostics have been primarily proposed to identify low-mass, Pop~III-dominated systems, their effectiveness in revealing a subdominant Pop~III component within high-$z$ galaxies -- where competing ionizing sources may significantly confuse the signal -- is unclear. 
In this work, we investigate the emission properties of a sample of massive halos hosting Pop~III star formation during the EoR, selected from the large-scale cosmological \texttt{dustyGadget} simulations presented in \citet{DiCesare_2023, Venditti_2023}. Our goal is to define robust criteria for identifying ongoing Pop~III activity even in complex, enriched systems, which have long remained uncharted territory within the landscape of Pop~III studies.

In \citet{Venditti_2023} we showed that similar ``hybrid'' Pop~III systems evolve during the EoR through a chaotic assembly resulting in a strongly inhomogeneous distributions of stars, gas, metals and dust. In addition, the short lifetimes of massive Pop~III stars imply a rapid fading of their ionizing radiative output \citep[e.g.,][]{Schaerer_2002, Schaerer_2003, Johnson_relic2009, Katz_2023}. In principle, to accurately recover the spectral-energy distribution (SED) of these galaxies, the full three-dimensional, time-dependent equation of radiative transfer should be solved in each galaxy. However, fully self-consistent radiation-hydrodynamic simulations \citep[e.g.,][]{Jeon_starburst2019, Katz_2022} are computationally expensive and therefore only feasible at relatively small scales or for few, zoomed objects. Post-processing cosmological simulations with photo-ionization models offers an efficient, scalable alternative to create statistics of galaxies with nebular line emission predictions.

Photo-ionization codes like \texttt{Cloudy} \citep{Ferland_1998, Ferland_2003, Ferland_2017, Chatzikos_2023}, typically assuming simplified one-dimensional geometries and equilibrium conditions, are particularly effective at predicting nebular emission from ionized gas in the immediate vicinity of young stellar populations. These codes can also be coupled with stellar population synthesis (SPS) models for the sources, which reproduce the intrinsic integrated emission of simple stellar populations \citep[e.g.][]{Conroy_2009}. A widely adopted strategy -- following the seminal studies of \citet{Kewley_2013}, \citet{Orsi_2014} and \citet{Shimizu_2016} -- combines outputs from cosmological hydrodynamical simulations \citep[e.g.][]{Hirschmann_2017, Hirschmann_2019, Pallottini_2019, Pallottini_2022} or semi-analytical models \citep[e.g.][]{Baugh_2022} with pre-computed \texttt{Cloudy} grids such as those presented by \citet{Gutkin_2016} and \citet{Feltre_2016}. Originally introduced by \citet{Charlot_Longhetti_2001}, this framework usually relies on effective, galaxy-averaged parameters to describe the collective emission from HII regions and diffuse gas ionized by successive generations of stars.

However, robust determinations of the ionization, thermal and chemical properties of galaxies require a multiphase gas model. Assuming uniform, low-density conditions in high-density environments has been proved to significantly bias UV and optical nebular diagnostics \citep{Martinez_2025}. Joint JWST and ALMA observations also provide direct evidence for a multi-zone evolution of electron densities at high redshifts, clearly indicating the need for a spatially resolved modeling of galaxy emission \citep{Harikane_2025}. Motivated by these considerations, here we follow a different approach than the aforementioned works, directly assigning SPS spectra to individual stellar populations within our simulated galaxies, and performing tailored \texttt{Cloudy} simulations for each HII region. A similar methodology has been successfully adopted in the past, e.g., for the \texttt{BlueTides} \citep{Wilkins_2020} and \texttt{Simba} simulations \citep{Garg_2022}.

The details of our emission model and of the adopted simulation suite, as well as the halo sample selected for this study, are expanded in Section~\ref{sec:methods}. Section~\ref{sec:results} presents our main results, with a particular focus on HeII (Section~\ref{sec:results_HeII}), HI (Section~\ref{sec:results_HI}) and [OIII] (Section~\ref{sec:results_OIII}) line emission from Pop~III stellar populations and confusing Pop~II sources. We discuss these results in Section~\ref{sec:discussion}, and finally summarize our conclusions in Section~\ref{sec:conclusions}.

\section{Methods}
\label{sec:methods}

\subsection{Simulation suite overview}
\label{sec:methods_dustyGadget_simulations}

The hydrodynamical code \texttt{dustyGadget} \citep{Graziani_2020} is an improved version of \texttt{Gadget-2/3} \citep{Springel_2005_Gadget-2, Springel_2008_Gadget-3} accounting for self-consistent dust production and evolution on top of the chemo-dynamical extensions of the original Smoothed Particle Hydrodynamics (SPH) scheme \citep{Tornatore_2007_chemicalFeedback, Maio_2007, Maio_2010, Maio_2011}. Our study is carried out on a suite of eight \texttt{dustyGadget} cosmological simulations (hereafter dubbed as U6 - U13), each with a size of $50h^{-1} ~\si{cMpc}$ and a dark matter (gas) particles mass resolution of $3.53 \times 10^7 h^{-1} ~ \si{M_\odot}$ ($5.56 \times 10^6 h^{-1} ~ \si{M_\odot})$.
All the cubes share common assumptions for the $\mathrm{\Lambda}$CDM cosmology\footnote{The adopted setup is consistent with \citet{Planck_2015}$: \Omega_{\mathrm{m,0}} = 0.3089$, $\Omega_{\mathrm{b,0}} = 0.0486$, $\Omega_{\mathrm{\Lambda},0} = 0.6911$ and $h = 0.6774$.}, and a common feedback setup based on \citet{Graziani_2020}, but starting from different cosmological initial conditions and thereby representing eight independent realizations of cosmic structure formation. The simulations are carried out from $z \approx 100$ down to $ z \approx 4$, while this work will focus on the redshift range $6.5 \lesssim z \lesssim 9$, as in \citet{Venditti_2023}.\\

Star formation is implemented in the two-phase inter-stellar medium (ISM) model introduced by  \citet{Springel_Hernquist_2003} and further extended in  \citet{Tornatore_2007_PopIII, Maio_2010, Graziani_2020}.
Two stellar populations (hereafter Pop~II and Pop~III stars) are followed, with a critical metallicity value of $Z_\mathrm{crit} = 10^{-4} ~ \si{Z_\odot}$ regulating their transition\footnote{A value of $Z_\mathrm{\odot} = 0.02$ is assumed for the solar metallicity  \citep{Anders_Grevesse_1989}.}. 
Collisionless particles representing stellar populations of $\sim 2 \times 10^6 ~ \si{M_\odot}$ are formed, with IMFs assigned according to their metallicity $Z_\star$:
\begin{enumerate}
    \item for Pop~II stellar particles ($Z_\star \geq Z_\mathrm{crit}$), a standard Salpeter IMF \citep{Salpeter_1955} is assumed in the range [0.1, 100] \si{M_\odot}. Mass and metallicity-dependent yields describing the metal pollution from long-lived, low-intermediate mass stars \citep{vanDenHoek_Groenewegen_1997}, high mass stars (> 8 \si{M_\odot}), dying as core-collapse SNe \citep{Woosley_Weaver_1995}, and TypeIa SNe \citep{Thielemann_2003}, are consistently associated with these populations;
    \item for Pop~III stars ($Z_\star < Z_\mathrm{crit}$), a Salpeter-like IMF in the mass range [100, 500] \si{M_\odot} is assumed, as well as mass-dependent yields describing the metal pollution from stars in the PISN range [140, 260] \si{M_\odot} \citep{Heger_Woosley_2002}. As most of the spectral-hardness diagnostics proposed for the identification of active Pop~III stellar populations at high redshifts rely on the prediction that Pop~III stars are predominantly massive, here we solely focus on top-heavy IMF models (also see Section~\ref{sec:methods_intrinsic_emission} for the role of the Pop~III IMF in our spectral model)\footnote{Note that the exact shape of the Pop~III IMF is still extremely uncertain, and stellar-archaeology studies show that a precise modeling of the low-mass end (extended down to $\sim 1 - 10 ~\Msun$, as demonstrated by the prevalence of faint SN signatures among Pop~III descendants) is required to reproduce the detailed nucleosynthetic pattern of old, metal-poor stars in the local Universe \citep[e.g.][]{deBennassuti_2014, deBennassuti_2017, Fraser_2017}. The impact of changing the shape and mass range of the IMF in a way that influences the power at high masses -- and therefore the HeII emission and the expected number of PISNe -- has been further explored in \citet{Venditti_2024_HeIIAnalyticalModel, Venditti_2024_PISNe}.}.
\end{enumerate}

Pop~III stars outside the PISN mass range and Pop~II/I stars with masses $\geq 40 ~ \si{M_\odot}$ are assumed to directly collapse into black holes (not followed explicitly in the simulations) and do not contribute to the chemical enrichment.
Finally, we conservatively assume that Pop~II stellar populations have a long tail of low-mass stars that can survive down to $z = 0$, while Pop~III stars, due to their high masses, are assumed to die after a relatively short time ($\approx 3$~Myr).

Note that the metallicity threshold below which star formation transitions towards a top-heavy-IMF typical of the first stars is believed to be primarily driven by an increase in gas metallicity ($\gtrsim 10^{-4} - 10^{-3} ~\Zsun$, \citealt{Omukai_2000, Bromm_2001_fragmentation, Bromm_Loeb_2003, Frebel_2007}). However a scenario with a dust-driven transition has also been suggested at lower critical metallicities ($\sim 10^{-6} - 10^{-4} ~\Zsun$, \citealt{Omukai_2005, Schneider_2012, Chiaki_2013}). Changing the value of the critical metallicity has a non-trivial effect on the rate of Pop~III star formation, as lower values can result in a larger number of newly formed stellar particles being classified as Pop~IIIs at any given time, while increased feedback from these top-heavy populations will result in turn in faster enrichment of the surrounding gas, potentially reducing their number at the following simulation snapshot. The impact of self-consistently changing $Z_\mathrm{crit}$ within the range $10^{-6} - 10^{-3} ~\Zsun$ has been explored for example by \citet{Maio_2010}, finding an overall $\sim 1$~dex scatter in the Pop III SFRD at $z \sim 11 - 15$, with lower $Z_\mathrm{crit}$ associated with lower SFRD at any given redshift in this range. 

The interested reader can refer to \citet{Graziani_2020} for further details on the \texttt{dustyGadget} code, while the adopted simulations are extensively discussed in \citet{DiCesare_2023} and \citet{Venditti_2023}. \citet{DiCesare_2023} placed our results for global scaling relations and for the assembly of stellar and dust mass across time in the context of available observations and other existing models. \citet{Venditti_2023} focused instead on the properties of galaxies hosting Pop~III star formation down to the EoR and their statistics.  

\subsection{Spectral modeling}

In this section we describe the steps of the \texttt{dustyGadget} post-processing pipeline reconstructing 
the spectra of simulated galaxies. We first determine the intrinsic spectrum of stellar populations simulated as stellar particles with assigned age ($t_\star$) and metallicity ($Z_\star$); see Sec. \ref{sec:methods_intrinsic_emission} for technical details on how the spectra are determined from Binary Population and Spectral Synthesis databases. As second step, the continua of particles with $t_\star \leq 10$~Myr are post-processed with \texttt{Cloudy} under specific assumptions on their surrounding ISM as described in Sec. \ref{sec:methods_NEL}. Finally, under the assumption that galaxies are spatially unresolved and their ISM is optically thin to nebular emission lines, we reconstruct the total spectrum by summing up the contributions of all stellar particles.

\subsubsection{Intrinsic spectra}
\label{sec:methods_intrinsic_emission}

To model the spectra emitted by Pop~II particles we adopt the \texttt{BPASS} SPS model \citep{Stanway_2016, Stanway_Elridge_2018}. \texttt{BPASS} covers a wide range of ages and metallicities (down to $5 \times 10^{-4} ~\Zsun$), also accounting for binary interactions between stars, which can induce mass loss or gain and affect stellar evolution, especially at low metallicity (e.g. figure~1 of \citealt{Stanway_Elridge_2018}), as well as Wolf-Rayet (WR) stars following the model of \citet{Sanders_2015}, as described in \citet{Eldridge_2017} (but see Section~\ref{sec:discussion_other_contaminants} for a discussion of other potential contaminants to the HeII emission). 
Here we also adopt a Salpeter IMF in the range [0.1, 100]~\si{M_\odot}, consistent with the \texttt{dustyGadget} simulations.
Intrinsic SEDs of Pop~III stellar populations are determined adopting the \texttt{Yggdrasil}\footnote{\url{https://www.astro.uu.se/~ez/yggdrasil/yggdrasil.html}} SPS model \citep{Zackrisson_2011}, assuming zero-metallicity populations formed in an instantaneous burst and following a Salpeter IMF in the range [50, 500]~\si{M_\odot}\footnote{This choice corresponds to the ``Pop~III.1'' case in the original paper, which is closest to the IMF adopted in our simulations. Note that an increase in the HeII emissivity by a factor $\approx 2.7$ would be found at the Zero-Age Main Sequence (ZAMS) by adopting our Salpeter-like IMF in the range $[100, 500] ~\Msun$ (from the model of \citealt{Schaerer_2002}). This discrepancy will decrease as the populations age and the massive component dies out.}.

The above models provide the intrinsic luminosity per unit wavelength ($L_\lambda$) as a function of $t_\star$ and $Z_\star$, normalized by their total stellar mass ($M_\mathrm{\star,ref} = 1 / 10^6 ~ \si{M_\odot}$ for \texttt{Yggdrasil}/\texttt{BPASS} spectra respectively). No \textit{a-priori} assumptions are made on the binary fraction of Pop\ II stars ($f_\mathrm{bin}$): we safely consider the two extreme cases in which Pop\ II are either composed of all binary stars ($f_\mathrm{bin} = 1$) or all single stars ($f_\mathrm{bin} = 0$); in this paper we will discuss the case of standard Pop~II stellar populations ($f_\mathrm{bin} = 0$), while the contribution of X-ray binaries to the HeII signal will be explored in a future work (Saggini et al. in preparation). \texttt{Yggdrasil} spectra, on the other hand, assume single stellar evolutionary tracks, as the binary fraction of Pop~III stars is still largely unconstrained \citep[but see][]{Stacy_binary2013, Liu_binary2021}\footnote{Note that mass exchanges in Pop~III and extremely-metal-poor binary stars have been found to enhance their total HI, HeI, HeII and LW photon yields by up to 46\%, 24\%, 2\%, and 89\%, respectively, because of the longer lifetime of the companion star \citep{Tsai_2023}.}.

\subsubsection{Nebular emission}
\label{sec:methods_NEL}

The intrinsic SEDs of stellar populations with $t_\star < 10$~Myr is given as input to \texttt{Cloudy}, under specific assumptions on the properties of the surrounding gas, as described below.

We assume a spherically symmetric, ionization-bounded nebula surrounding the stars, with constant H number density ($n_\mathrm{H}$), and case-B recombination. The inner radius of the cloud is $r_\mathrm{in} = 0.1$~pc, and the photoionization calculations are stopped when the temperature falls below 100~K or the electron density falls below 1\% of $n_\mathrm{H}$, to ensure that the Str{\"o}mgren sphere is not exceeded by far (as in \citealt{Charlot_Longhetti_2001}).

We also compute the ionization parameter ($U$) corresponding to each particle configuration. $U$ is defined as the ratio of the rate of ionizing photons to gas density at distance $r$ from the ionizing source:
\begin{equation}
    U(r,  Z_\star, t_\star) = \frac{Q (Z_\star, t_\star)}{4 \pi r^2 n_\mathrm{H} c},
    \label{eq:ionisation_parameter}
\end{equation}
\noindent
where the number of ionizing photons per unit time $Q(Z_\star, t_\star)$ is computed as:
\begin{equation}
    Q (Z_\star, t_\star) = \frac{1}{hc} \frac{M_\star}{M_\mathrm{\star,ref}} \int_0^{\lambda_\mathrm{L}} \lambda L_\mathrm{\lambda} (Z_\star, t_\star) \dd \lambda.
\end{equation}
\noindent 
In the above formula $\lambda_\mathrm{L} = 912$~\si{\angstrom}, $M_\star / M_\mathrm{\star,ref}$ is the ratio of the total stellar mass of the stellar population to the normalization mass of \texttt{BPASS}/\texttt{Yggdrasil}; $h$ and $c$ are respectively the Planck constant and the speed of light. The ionization parameter computed at $r_\mathrm{in}$, provides the appropriate normalization of the input spectrum. 

Due to our limited resolution, as in \citet{Venditti_2024_HeIIAnalyticalModel, Venditti_2024_PISNe}, we introduce an efficiency factor $\eta_\mathrm{III} < 1$ defined as:
\begin{equation}
    M_\mathrm{III} = \eta_\mathrm{III} M_\mathrm{III,res},
    \label{eq:popIIIResolutionElement}
\end{equation}
to account for an ``effective'' Pop~III stellar mass ($M_\mathrm{III}$), which can be lower than the mass of the associated stellar particle ($M_\mathrm{III,res} \sim 2 \times 10^6 ~ \si{M_\odot}$, Section~\ref{sec:methods_dustyGadget_simulations}). $M_\mathrm{III,res}$ can be interpreted as the amount of extremely metal-poor gas that is available for star formation, while $M_\mathrm{III}$ is the amount of stellar mass actually produced in a single star-formation event; therefore, $\eta_\mathrm{III}$ parametrizes the potential impact of a reduced star-formation efficiency on our emission model. As our goal is to identify extreme Pop~III stellar populations that may be distinguishable even within an overall enriched environment, here we assume an optimistic reference value of $\eta_\mathrm{III} = 0.3$ (unless otherwise stated), consistent with simulations of Pop~III star formation in atomic-cooling halos at late times (e.g. \citealt{Greif_Bromm_2006, Greif_2008, Bromm_2009, Sugimura_2024, Jeong_2026}, also see our discussion in Section~\ref{sec:discussion_inhomogeneous_enrichment}), and with the inferred Pop~III mass of the ``hybrid'' Pop~III candidate RXJ2129-z8HeII \citep{Wang_2024}, and of the HeII emitter identified in the environment of GN-z11 (which has been proposed to be explained by a Pop~III mass of $\sim 2 \times 10^4 - 6 \times 10^5 ~\Msun$, \citealt{Maiolino_2026, Ubler_2026, Rusta_2026}); note that even larger Pop~III masses, up to $\sim 10^6 ~\Msun$, have been suggested for Pop~III starbursts within atomic-cooling halos immersed in a strong LW background of $\sim 10^4 ~\si{J_{21}}$ \citep{Jeong_2026}. 

This rescaling is only applied in post-processing to achieve a more realistic modeling of the emission of Pop~III stellar populations, in line with higher-resolution simulations and observed candidates. On the other hand, it is not reflected in the numerically tracked feedback. While the full implications of any changes in the Pop III SFE and IMF need to be explored self-consistently, here we note that having larger total Pop~III masses (and/or a larger relative amount of massive Pop~III stars in the populations) is expected to increase feedback from the stellar particles \citep[e.g.][]{Cueto_2024, Hutter_2025}, likely reducing the overall pristine reservoir over time. Therefore, the SFE and IMF adopted in the simulations can be considered a conservative choice for tracking the persistence of pristine gas.

As the adopted hydrodynamical simulations do not have sufficient mass resolution to accurately describe the gas properties in individual HII regions, \textit{a-priori} assumptions on the physical state of the gas are necessary. Different values of the hydrogen number density ($n_\mathrm{H}$), gas metallicity ($Z_{\rm gas}$) and dust-to-metal ratio ($\xi_\mathrm{dust}$) have been selected to reproduce typical optical-line diagnostics (see e.g. Section~\ref{sec:results_OIII}) in observed high-$z$ galaxies (Graziani et al. in preparation). For consistency with the \texttt{Yggdrasil} spectral database, pristine gas is assumed around Pop~III stellar populations, while the main assumptions on the clouds surrounding Pop~II stellar populations are summarized in the paragraphs below.

\paragraph{Gas number density}
Hydrogen number densities from $n_\mathrm{H} = 10^3 ~\si{cm^{-3}}$ to $n_\mathrm{H} = 10^4 ~\si{cm^{-3}}$ are assigned as a step function of the age of the enclosed stellar particles, according to four main fiducial configurations (named ``naked'', ``semi-naked'', ``semi-embedded'' and ``embedded''), which describe progressively higher gas densities around young stellar populations, as summarized in Table~\ref{tab:number_density_config}. 
Note that, with this setup, only two values of $n_\mathrm{H}$ are possible for the gas surrounding Pop~III particles, which are always younger than 3~Myr with our assumed IMF (see Section~\ref{sec:methods_dustyGadget_simulations}): $n_\mathrm{H} = 10^{3.7} ~\si{cm^{-3}}$ and $n_\mathrm{H} = 10^4 ~\si{cm^{-3}}$.\\
We assume that young stellar particles will be surrounded by dense/cold gas, close to the cloud originating the Jeans collapse, while older stars  live in regions with densities closer to the dense/diffuse ISM (following observations in our Galaxy, e.g. \citealt{Krumholz_2017}). Due to the limits imposed by our mass resolution, we cannot however resolve the real densities of cold molecular clouds. As a reference, the highest resolved ISM densities, when SPH gas particles are projected into a 512$^3$-cell cartesian grid capturing our galaxies at kpc scale, are $\sim 100 ~\si{cm^{-3}}$ -- a value typically assumed in early photoionization modeling \citep[e.g.][]{Nakajima_2023}. We then explore values at least two orders of magnitudes higher, corresponding to min./max. typical values from recent literature at high redshifts \citep[e.g.][]{Isobe_2023, Yanagisawa_2024, Harikane_2025}.\\
These modeling assumptions, and the comparison with published line-ratio observed diagnostics (e.g. O32, R23) will be detailed in Graziani et al. in preparation.

\begin{table}
    \centering
    \resizebox{\columnwidth}{!}{
        \begin{tblr}{
            colspec={|r|c|c|c|c|c|},
            cell{1}{1} = {r=2}{},
            cell{1}{6} = {r=2}{},
            cell{1}{2} = {c=4}{c},
            cell{3}{6} = {r=4}{c},
            row{1-2} = {ht=4mm},
            row{3-6} = {ht=4mm},
            cell{3-6}{1} = {font=\small},
            vspan=even,
            hlines,
            vlines
        }
            & $t_\star$ [Myr] & & & & \\
            & $[0-0.5]$ & $[0.5-1]$ & $[1-3]$ & $[3-10]$ & \\
            \textbf{naked}
                & \SetCell{bg=gray!35} 3.7
                & \SetCell{bg=gray!35} 3.7
                & \SetCell{bg=gray!35} 3.7
                & \SetCell{bg=gray!15} 3
                & \rotatebox{270}{\footnotesize Log$(n_\mathrm{H}/\si{cm^{-3}})$} \\
            \textbf{semi-naked}
                & \SetCell{bg=gray!60} 4
                & \SetCell{bg=gray!35} 3.7
                & \SetCell{bg=gray!35} 3.7
                & \SetCell{bg=gray!15} 3
                & \\
            \textbf{semi-embedded}
                & \SetCell{bg=gray!60} 4
                & \SetCell{bg=gray!60} 4
                & \SetCell{bg=gray!35} 3.7
                & \SetCell{bg=gray!15} 3
                & \\
            \textbf{embedded}
                & \SetCell{bg=gray!60} 4
                & \SetCell{bg=gray!60} 4
                & \SetCell{bg=gray!60} 4
                & \SetCell{bg=gray!15} 3
                & \\
        \end{tblr}
    }
    \caption{Summary of the four fiducial configurations in our \texttt{Cloudy} runs for the H number density of the clouds ($n_\mathrm{H}$) as a function of the age of their enclosed stellar population ($t_\star$). Each column reports the value of the number density -- expressed as Log$(n_\mathrm{H}/ \si{cm^{-3}})$ -- assigned to a given range of ages in the various configurations (one per table row, with the name of each configuration in the first column).}
    \label{tab:number_density_config}
\end{table}

\paragraph{Gas-phase metallicity and abundance ratios}
The gas-phase metallicity ($Z_\mathrm{gas}$) of the nebula around Pop~II particles is assumed to linearly scale with $Z_\star$ through a fixed boosting factor $\alpha$. For young stellar populations the typical assumption $Z_\mathrm{gas} = Z_\star$ is certainly acceptable, as the stars follow the chemical composition of the parent cloud and have not moved far away from their birth environment. However, this assumption breaks for older stellar populations and for the more chemically evolved ISM of galaxies in the upper end of the galaxy main sequence; recent studies report an average $\alpha \sim2.5$ enhancement in the star-forming regions of $z \sim 3$ galaxies, with a scatter in the range $1 \leq \alpha \leq 5$ \citep{Cullen_2021}. Here we assume a fiducial value of $Z_\mathrm{gas} = 2.5 \times Z_\star$, while additional cases at extreme $\alpha$-values will be explored by Graziani et al. in preparation.\\
As the adopted simulations only explicitly track six atomic elements, we follow  \citet{Gutkin_2016} to set individual abundances. In particular, for $Z_\mathrm{gas} = Z_\odot$ we adopt the total abundances ($A_{i,\mathrm{tot}}$) from their Table~1\footnote{Note that the total solar abundances from \citet{Gutkin_2016} are given at $\Zsun = 0.01524$, and therefore we rescale them to match our assumed solar metallicity ($\Zsun = 0.02$, \citealt{Anders_Grevesse_1989}).}. Non-solar metallicity cases derive metal abundances by rescaling with the total metallicity of the cloud $Z_\mathrm{tot} \equiv M_\mathrm{Z} / M_\mathrm{gas} = Z_\mathrm{gas} / (1 - \xi_\mathrm{dust})$, where $M_\mathrm{gas}$ is the total gas mass of the cloud, and $M_\mathrm{Z}$ the total metal mass combining metals in both gas and dust ($M_\mathrm{dust}$) phase; $\xi_\mathrm{dust} \equiv M_\mathrm{dust} / M_\mathrm{Z}$ defines the dust-to-metal mass ratio parameterizing the total amount of metals depleted onto dust grains. \\
As final note, we also implemented the corrections considered by 
\citet{Gutkin_2016} on N and C to take into account the metallicity dependence of their secondary-nucleosynthesis component, following the indications detailed in the original paper.

\paragraph{Dust grains}
Although \texttt{dustyGadget} simulates the formation and evolution of cosmic dust in the ISM of galaxies, dust grains in unresolved HII regions require additional modeling, as is the case for the gas properties. \\
The depletion of atomic metals onto dust grains is also modeled following \citet{Gutkin_2016}. For dust-to-metal ratios close to the solar value ($\xi_\mathrm{dust} = \si{\xi_{dust,\odot}} = 0.36$), we adopt the depletion factors ($d_{i}$) in their table~1\footnote{These indicate the fraction of $n_{i,\mathrm{tot}}$ that remains in gas phase after the formation of dust ($d_i = 1 - f_{\mathrm{dpl},i}$, with $f_{\mathrm{dpl},i}$ the fraction of element $i$ depleted onto dust grains).}. 
Depletion factors associated with non-solar dust-to-metal ratios are obtained with a linear interpolation of the depletion factors  associated with $\xi_\mathrm{dust} = 0$, \si{\xi_{dust,\odot}} and 1, ensuring that $f_{\mathrm{dpl},i}= 0/1$ for $\xi_\mathrm{dust} = 0/1$ respectively\footnote{Note that the value $\xi_\mathrm{dust} = 1$ is unphysical, as it would require all metals to be locked in grains (including non-refractory elements); it is therefore considered only as a limit case.}, and that solar depletion factors are recovered for solar dust-to-metal ratio. With the above values, gas-phase abundances are simply computed as $A_{i,\mathrm{gas}} = d_i A_{i,\mathrm{tot}}$ once metal depletion into grains is accounted for. In the present setup, we assume that HII regions are dust-poor and safely assume a low value of $\xi_\mathrm{dust} = 3 \% ~\si{\xi_{dust,\odot}}$.\\ 
While Small Magellanic Cloud-like \citep[e.g.][]{Mancini_2016} or SN attenuation curves \citep[e.g.][]{McKinney_2025} have been shown to better reproduce the observed photometry of $z > 6$ galaxies, other dust physical properties are borrowed from Milky Way dust implemented in \texttt{Cloudy}. The default ISM grain abundances are derived from the dust-to-gas mass ratio ($\mathcal{D}$) corresponding to our assumed $Z_\mathrm{tot}$ and $\xi_\mathrm{dust}$, so that $\mathcal{D} = \xi_\mathrm{dust} Z_\mathrm{tot}$. Grain size distributions and optical properties are from \cite{Mathis_1977} and \cite{Martini_1991}.\\
As final remark, we note that most of the attenuation is expected to arise from dust within the galaxy ISM, rather than from the dust within the HII region (see e.g. sec. 3.4 of \citealt{Venditti_2023} for a discussion of the dust column density to our Pop~III sources across typical lines-of-sight). An accurate determination of dust attenuation is deferred to a future work.

\subsection{Our halo sample}
\label{sec:methods_haloSample}

\begin{figure}
    \centering
    \includegraphics[width=\linewidth]{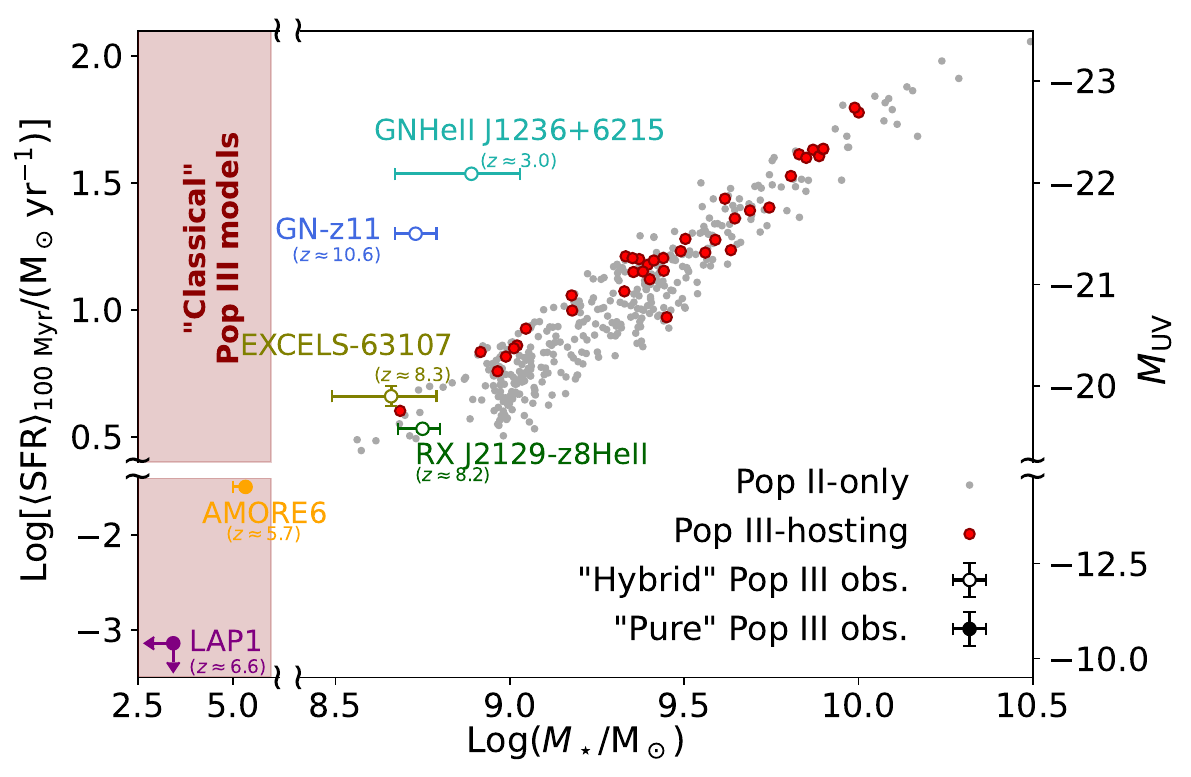}
    \caption{Total (Pop~III + Pop~II) SFR averaged over 100 Myr ($\langle \mathrm{SFR} \rangle_\mathrm{100~Myr}$) as a function of total stellar mass ($M_\star$) for the Pop~III-hosting halos considered in this study (\textit{red circles}), compared with benchmark Pop~II-only halos (\textit{grey dots}).
    The right axis shows the corresponding UV magnitudes ($M_\mathrm{UV}$) assuming a simple SFR-UV luminosity conversion as in \citet{Madau_Dickinson_2014} (see text).
    We compare with the $M_\mathrm{UV}$ vs $M_\star$ of faint, ``pure'' Pop~III galaxy candidates (LAP1, \citealt{Nakajima_2025}, and AMORE6, \citealt{Morishita_2025}, \textit{filled, colored circles}), and with brighter, ``hybrid'' Pop~III candidates (RXJ2129-z8HeII, \citealt{Wang_2024}, EXCELS-63107 \citealt{Cullen_2025}, GN-z11, \citealt{Bunker_2023, Maiolino_2024, Maiolino_2026, Ubler_2026}, and GNHeII J1236+6215, \citealt{Mondal_2025}, \textit{empty, colored circles}) proposed at $z \gtrsim 3$. 
    Observed candidates cover a broad range of $M_\star$ and $M_\mathrm{UV}$, with our halo sample lying in a regime largely unexplored by ``classical'' Pop~III models: while classical models typically focus on low-mass, Pop~III-dominated systems, here we consider massive halos dominated by Pop~II star formation, but hosting sub-dominant Pop~III-forming clumps in their environment. Note that both axes have a broken range marked by ``$\approx$'' symbols, in order to fit the faint LAP1 and AMORE6 constraints on the bottom-left corner.}
    \label{fig:MS}
\end{figure}

The sample of Pop~III hosts considered in this study is shown in Figure~\ref{fig:MS}, in comparison with typical main-sequence Pop~II-only halos in a similar mass range for which we also modeled the nebular emission, and which will be used as a benchmark in the present work; the physical properties of these Pop~III systems are also summarized in Table~\ref{tab:PopIII_halo_sample} of Appendix~\ref{sec:app_PopIII_halo_properties}. To enable a rough comparison with Pop~III galaxy candidates observed at high redshifts, we adopted the simple calibration proposed by \citet{Madau_Dickinson_2014} to derive the UV luminosity of galaxies with given total SFR (here averaged over a 100 Myr timescale, $\langle \mathrm{SFR} \rangle_\mathrm{100~Myr}$), assuming a conversion factor\footnote{Note that, while both the stellar mass and SFR -- and therefore the overall emission -- of the galaxies considered in this study is largely dominated by Pop~IIs, the UV light produced by stellar populations in these high-$z$ galaxy environments may differ substantially from typical local galaxies, as it will depend on specific characteristics of the stellar populations such as their metallicity and IMF, as well as their star-formation history. Therefore, the comparison provided here is to be intended as only indicative.} $\kappa_\mathrm{UV} = 1.15 \times 10^{-28} ~\si{\Msun.yr^{-1}.erg^{-1}.s.Hz}$.  

We compare with the UV magnitudes and inferred stellar masses of Pop~III galaxy candidates observed at $z \gtrsim 3$, including systems with no detected metal lines, which could be explained by a single Pop~III stellar population in the absence of metal enrichment (``pure'' Pop~III systems, i.e. LAP1, \citealt{Nakajima_2025}, and AMORE6, \citealt{Morishita_2025}), and systems in which a Pop~III component has been suggested to explain part of the observed emission, even in the presence of metal lines (``hybrid'' Pop~III systems, i.e. RXJ2129-z8HeII, \citealt{Wang_2024}, EXCELS-63107, \citealt{Cullen_2025}, GN-z11, \citealt{Bunker_2023, Maiolino_2024, Maiolino_2026, Ubler_2026}, and GNHeII J1236+6215, \citealt{Mondal_2025}). 

The key point of innovation of this work lies in our specific focus on massive systems ($M_\mathrm{vir} \gtrsim 10^{11} ~\Msun$, $M_\star \gtrsim 10^9 ~\Msun$), a population that has remained largely unexplored in the context of Pop~III formation. Most theoretical and observational efforts have in fact targeted the dominant population of faint, low-mass hosts (consistent with ``pure'' Pop~III galaxy candidates forming less than $\lesssim 10^6 ~\Msun$ in stellar mass, e.g. \citealt{Nakajima_2025, Morishita_2025}). Resolving such small star-forming regions requires extremely high numerical resolutions to capture the smallest structures. However, simulations performed at these resolutions are typically limited to small cosmological volumes (a few cMpc per side or less), preventing a comprehensive sampling of rare, massive galaxies that may still host detectable Pop~III components (as in the ``hybrid'' Pop~III galaxy candidates proposed in recent literature). The large \texttt{dustyGadget} boxes overcome this limitation, enabling the first systematic investigation of massive Pop~III-hosting halos across cosmic time \citep{Venditti_2023}.

We note that, rather than attempting to give a final answer on the physical plausibility of similar Pop~III systems, the present study aims to address two main questions: (i) if Pop~III stars can survive around massive galaxies, would they be observable? (ii) Would we be able to tell them apart from nearby, second-generation stars? Broader questions regarding the conditions that enable the survival of metal-free gas, how late we can push the theoretical boundary for the formation of metal-free stars, and what are the most favorable environments for their search -- as well as the dependence of these predictions on the underlying physical model --, will be deferred to future studies. 
Particularly, pre-enrichment of massive halos and their associated pristine pockets, caused by an unresolved population of minihalos at earlier epochs, may reduce the number of these systems that can preserve their metal-free nature at late cosmic epochs (see e.g. \citealt{Trenti_Stiavelli_2009, Johnson_2010, Regan_2020, Hicks_2021, Hicks_2024}). While a robust account of this effect on the statistics of late Pop~III hosts (presented in \citealt{Venditti_2023}) will require more in-depth studies, connecting between scales and thoroughly investigating the mechanisms that propagate metals at all scales, here we aim to answer a more fundamental question  regarding their observability, which will serve to motivate such further explorations. Nonetheless, we note that similar conclusions on the observability of massive Pop III hosts may apply even if their formation is confined to earlier epochs than our current forecasts, as the HeII emission properties of these hosts do not show significant evolution over the explored redshfit range.
The interested reader may refer to Section~\ref{sec:discussion_inhomogeneous_enrichment} for a qualitative discussion of the potential of Pop~III searches around massive galaxies for constraining inhomogeneous enrichment in overdense environments, and ultimately providing a crucial test on the overall progress of metal enrichment across space and time.

\section{Results}
\label{sec:results}

In this section we present our predictions for the nebular line emission arising from massive Pop~III-hosting halos. 
We begin by assessing the detectability of Pop~III stars in these systems using common spectral-hardness diagnostics, focusing on HeII (Section~\ref{sec:results_HeII}) and HI (Section~\ref{sec:results_HI}) line luminosities, and on the potential confusion arising from standard Pop~II populations. 
We then conclude by presenting the metal-line emission produced by enriched stellar populations in the same halos, to test the common assumption that these systems lie predominantly in a low-metallicity regime (Section~\ref{sec:results_OIII}).

We note that here we maintain a specific focus on nebular-emission-line diagnostics, while leaving the discussion of the continuum (including other proposed spectral features of interest for Pop~III populations such as Balmer breaks, 2-photon emission and UV $\beta$-slopes) to a future work. Therefore, we will only explicitly discuss the line luminosity emerging from the continuum (directly provided as a \texttt{Cloudy} output) rather than equivalent widths (EWs), as commonly done in the literature. This is motivated by two arguments of both practical and methodological nature: (i) for several observed candidates \citep[e.g.][]{Nakajima_2025, Maiolino_2024, Maiolino_2026, Ubler_2026}, the continuum arising from the underlying stellar population has remained elusive, making EW estimates inherently more uncertain than direct line-flux measurements; 
(ii) we therefore focus on line luminosities, which represent a ``clean'' \texttt{Cloudy} output and do not require additional assumptions about the continuum contribution from $t_\star \geq 10$~Myr stellar populations, for which we only modeled intrinsic spectra without applying nebular corrections.

Furthermore, we note that all results presented in this paper have been collapsed in redshift, since we find no significant evolution in the HeII line emission properties of Pop~III-hosting halos across $6.5 \lesssim z \lesssim 9$. The redshift evolution of other optical emission-line diagnostics -- less dependent on the Pop~III component -- will be discussed in Graziani et al. in preparation.

\subsection{HeII line emission}
\label{sec:results_HeII}

\subsubsection{Pop~III-forming clumps and their detectability}
\label{sec:results_HeII_PopIIIClumps}

\begin{figure*}
    \centering
    \includegraphics[width=\linewidth]{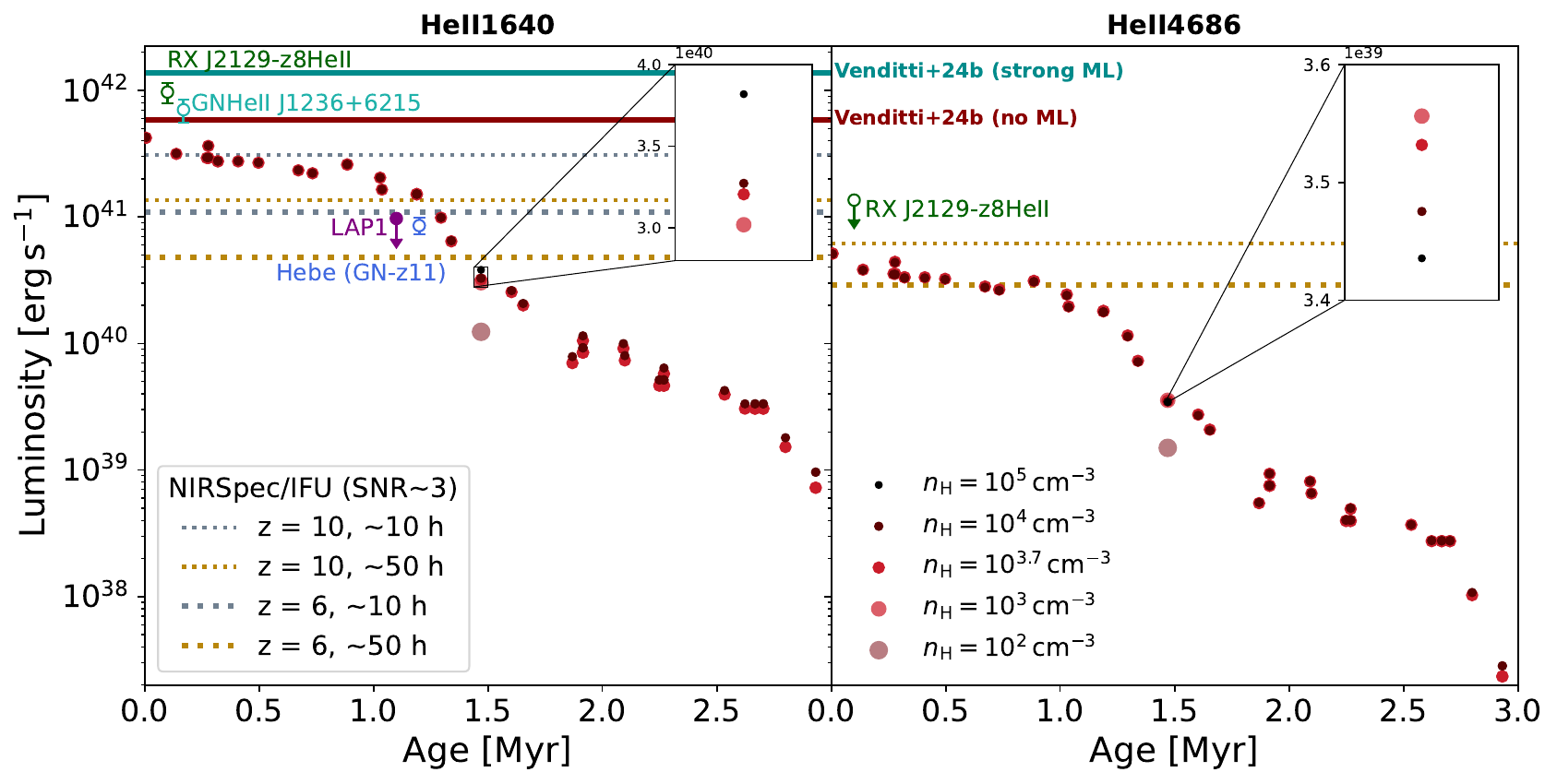}
    \caption{HeII emission line luminosities from individual Pop~III-forming clumps at 1640~\AA ~(\textbf{left}) and 4686~\AA ~(\textbf{right}), as a function of the age of the enclosed stellar populations, assuming a H number density around the stellar populations of $n_\mathrm{H} = 10^{4} ~\si{cm^{-3}}$ (\textit{dark-red, smaller dots}) and $n_\mathrm{H} = 10^{3.7} ~\si{cm^{-3}}$ (\textit{red, larger dots}) respectively, and an efficiency factor $\eta_\mathrm{III} = 0.3$ (corresponding to a Pop~III mass of approximately $6 \times 10^5 ~\Msun$). For a stellar population of age $\sim 1.5$~Myr, other values of $n_\mathrm{H}$ explored in the model are shown with progressively \textit{lighter shades of red} and progressively \textit{larger dots} (with a zoom around the blended points in the \textbf{top-right insets}). \textit{Dark-red/cyan horizontal, solid lines} show a comparison with the simple analytical model of \citet{Venditti_2024_HeIIAnalyticalModel}, assuming no/strong mass loss, and the same $\eta_\mathrm{III} = 0.3$. 
    Sensitivity thresholds at a $\mathrm{SNR} \sim 3$ for medium-resolution observations with JWST/NIRSpec/IFU are also shown as \textit{dotted lines}, considering $\sim 10$~h (\textit{grey}) and $\sim 50$~h (\textit{gold}) of total integration time for sources at $z = 10$ (\textit{thin}) and $z = 6$ (\textit{thick}). The HeII line luminosity constraints of the observed Pop~III candidates RXJ2129-z8HeII \citep{Wang_2024}, GNHeII-J1236+621 \citep{Mondal_2025}, LAP1 \citep{Nakajima_2025}, and Hebe \citep{Maiolino_2026} are also shown as \textit{colored circles}, with arbitrary positioning on the $x$ axis. The HeII emission decreases by up to $\sim 3$~dex with stellar age within the first 3~Myr. Only populations younger than $\sim 1$~Myr remain marginally detectable with $\sim 10$~h of observations up to $z \approx 10$ (HeII1640), or $\sim 50$~h up to $z \approx 6$ (HeII4686). In contrast, a much smaller variation of $\sim 0.5$~dex is found when varying $n_\mathrm{H}$ by 3 orders of magnitude.}
    \label{fig:HeIILuminosities_PopIIIClumps}
\end{figure*}

Figure~\ref{fig:HeIILuminosities_PopIIIClumps} shows a plot of the HeII1640 and HeII4686 line luminosities arising from individual Pop~III clumps as a function of the age of the enclosed stellar populations, assuming our reference number densities $n_\mathrm{H} = 10^{4} ~\si{cm^{-3}}$ and $n_\mathrm{H} = 10^{3.7} ~\si{cm^{-3}}$ for the gas surrounding the stars (Section~\ref{sec:methods_NEL}).

The main result is the strong decreasing trend of the HeII line luminosity with stellar age, as the massive stars producing most of the HeII-ionizing photons progressively die out; this is in agreement with previous studies (e.g., \citealt{Schaerer_2002, Schaerer_2003, Katz_2023}).
In \citet{Venditti_2024_HeIIAnalyticalModel}, we adopted a simple analytical model for the HeII1640 line luminosity produced by Pop~III clusters of given stellar mass, by simply convolving the HeII emissivity as a function of the mass of individual Pop~III stars from \citet{Schaerer_2002} with our assumed Pop~III IMF, under the assumption that the mass lost during our short Pop~III lifetimes is negligible. By comparing these analytical predictions with the results of our precise photoionization calculations, it appears evident that this is not a good approximation, as the HeII luminosity significantly drops even over a $\sim 3$~Myr timescale, and in fact these predictions always lie above our simulated data points.

To determine the observability of these systems, we also compute sensitivity limits with a signal-to-noise ratio (SNR) $\sim 3$ for $\sim$10 and $\sim$50 hours of observation at $z = 6$ and $z = 10$ with the JWST/NIRSpec spectrograph, using the JWST Exposure Time Calculator (ETC)\footnote{\url{https://jwst.etc.stsci.edu/}}. We adopt the same setup for the ETC as in figure~2 of \citet{Venditti_2024_HeIIAnalyticalModel}, considering the most favorable scenario for detectability derived in that work, i.e. the case of NIRSpec/IFU observations with medium resolution\footnote{Note that, in addition to an improved sensitivity with respect to Prism spectra, using the medium-resolution grating would also allow deblending of the HeII1640 line from the OIII]1661-1666 doublet.}, assuming a narrow line width of $50 ~\si{km.s^{-1}}$. Only Pop~III populations with an age below $\sim 1$~Myr are always marginally detectable in the considered redshift range\footnote{However, note that taking the continuum into account may slightly improve our odds, by boosting the observed line flux by an amount equivalent to the continuum level at the line wavelength, see e.g. the discussion of \citet{Venditti_2024_HeIIAnalyticalModel}.}, while pushing the integration time to $\sim 50$~hours only increases the range of detectability up to $\sim 1.5$~Myr at $z \sim 6$ for the HeII1640 line. The weak redshift dependence of our results appears to favor a search for the Pop~III pockets in the lower-$z$ bracket of our considered redshift range. 
Also note that a fivefold increase in exposure time improves sensitivity by only a factor $\sim 2$, implying diminishing returns for very deep pointings on a limited number of targets. A more effective strategy may therefore be to carry out wide, shallower searches for young, massive Pop~III systems across a large statistics of massive-galaxy environments, thereby maximizing our chances of capturing this rare but bright population. Moreover, large fields-of-view (FoVs) are needed to include potential systems at the periphery of the dark matter halos (see Section~\ref{sec:results_spatiallyResolvedObservations}), further supporting prioritizing area coverage over depth in small fields.

As the chosen gas number density determines the amount of gas available for HeII recombinations, we studied the impact of this assumption by exploring more extreme scenarios (from $10^2 ~\si{cm^{-3}}$ to $10^5 ~\si{cm^{-3}}$) for a Pop~III particle at the edge of detectability (with an age of $\sim 1.5$~Myr). We find a maximum span of $\sim 0.5$~dex in HeII luminosity over this wide range of values, much lower than the $\sim 3$~dex variability due to stellar age alone. 
A much larger impact comes from changing the amount of Pop~III mass produced in a single star-formation episode (parametrized by $\eta_\mathrm{III}$), which yields an approximately linear variation in the resulting HeII luminosity. This implies that, in practice, even young Pop~III stellar populations would only be observable in cases of particularly efficient star formation -- at least $\sim 10$ times more efficient than typical mini-halos ($\eta_\mathrm{III} \sim 0.01$). Indeed, if some portion of the gas within massive halos remains unpolluted, it may exhibit favorable conditions for hosting large Pop~III starbursts, making large $\eta_\mathrm{III}$ in these environments not only possible, but likely  (see the discussion in Section~\ref{sec:discussion_inhomogeneous_enrichment}).

\subsubsection{Integrated halo emission and confusion from standard Pop~II stellar populations}
\label{sec:results_HeII_PopIIIHalos}

\begin{figure*}
    \centering
    \includegraphics[width=\linewidth]{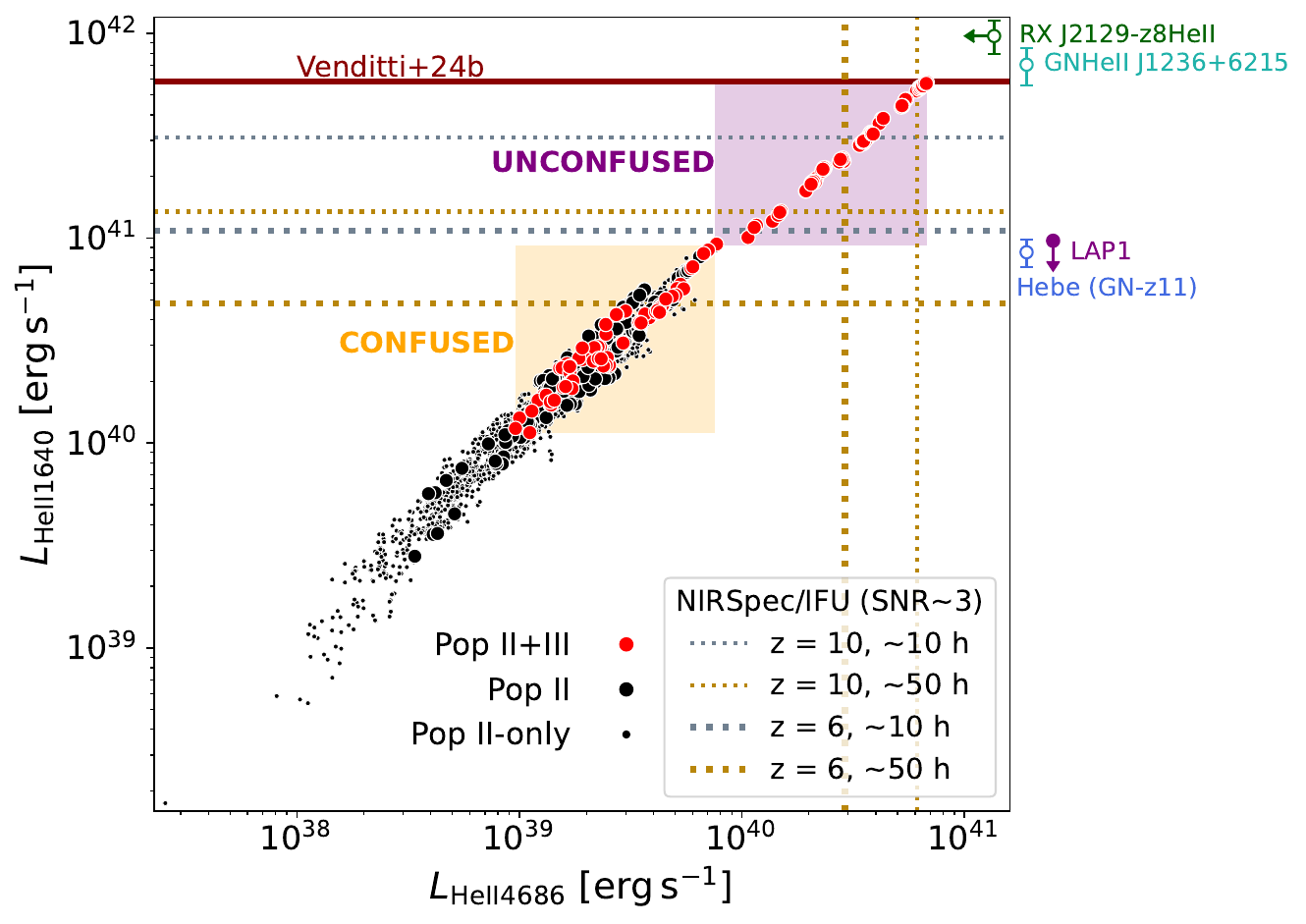}
    \caption{Integrated HeII line luminosity at 1640~\AA ~($L_\mathrm{HeII1640}$) vs 4686~\AA ~($L_\mathrm{HeII4686}$) arising from each halo within our Pop~III halo sample and from benchmark Pop~II-only halos at $z \approx 6.5 - 9$ (Section~\ref{sec:methods_haloSample}), assuming our reference \texttt{Cloudy} configurations (Section~\ref{sec:methods_NEL}). 
    The total (Pop~III + Pop~II) HeII emission arising from Pop~III-hosting halos and the contribution to the emission coming from confusing Pop~II populations in these halos is shown by \textit{red/black circles}, respectively; the HeII emission of benchmark Pop~II-only halos is shown instead by \textit{black dots}. 
    We identify two main regions of interest for Pop~III-hosting halos: a ``confused'' region (\textit{orange shaded area}), in which the HeII emission arising from Pop~III stars cannot be disentangled from a potential underlying Pop~II floor, and an ``unconfused'' region (\textit{purple shaded area}), which can only be reached through a Pop~III contribution, allowing a more straightforward identification. 
    The HeII luminosity of observed Pop~III candidates (RXJ2129-z8HeII, \citealt{Wang_2024}, GNHeII-J1236+6215, \citealt{Mondal_2025}, Hebe, \citealt{Maiolino_2026}, and LAP1, \citealt{Nakajima_2025}) and sensitivity thresholds for JWST/NIRSpec are shown as in Figure~\ref{fig:HeIILuminosities_PopIIIClumps}, as well as a comparison with the simple analytical model of fig.~2 of \citet{Venditti_2024_HeIIAnalyticalModel}. Note that for GNHeII-J1236+6215, Hebe and LAP1, no constraints are available on the HeII4686 line, therefore measures of the HeII1640 line luminosity are shown out of the $x$-axis boundaries of the plot to avoid confusion.}
    \label{fig:HeIILuminosities_haloSum}
\end{figure*}

As all the Pop~III-forming clumps analyzed in Section~\ref{sec:results_HeII_PopIIIClumps} lie within Pop~II-dominated halos, we must determine the level of contamination of the HeII signal that can arise from standard Pop~II stellar populations in the surrounding environment. In fact, most of these clumps would be found close enough to Pop~II clusters to be unresolved with the typical spatial resolution of NIRSpec/IFU (see e.g. our discussion in Section~\ref{sec:results_spatiallyResolvedObservations}).

In order to determine the maximum confusion expected from nearby Pop~IIs, in Figure~\ref{fig:HeIILuminosities_haloSum} we show the integrated HeII line emission arising from our sample of massive Pop~III hosts at $z \sim 6.5 - 9$ (Section~\ref{sec:methods_haloSample}) in our reference Cloudy configurations (Section~\ref{sec:methods_NEL}), computed by summing the emission from the nebulae surrounding all stellar populations within each halo, taking into account the individual age and metallicity of individual stellar particles at the corresponding simulation snapshot. The total signal produced by combined Pop~III and Pop~II stellar populations is compared to the contribution coming from Pop~II populations alone, and from a selection of control Pop~II-only halos. 
Our predictions for Pop~III-hosting halos span a wide range of luminosities ($\sim 1.7$~dex), and a large portion of this scatter lies in the region of the plane in which the Pop~III emission is completely confused by standard Pop~II populations. However, the brightest Pop~III systems (above $L_\mathrm{HeII1640} \sim 10^{41} ~\si{erg.s^{-1}}$ / $L_\mathrm{HeII4686} \sim 10^{40} ~\si{erg.s^{-1}}$) would be reliably identified against these Pop~II contaminants. 
Although our simulated galaxies only reach SFRs up to $\sim 10^2 ~\si{\Msun.yr^{-1}}$, by extrapolating the relation between the HeII luminosity produced by standard Pop~II populations and the SFR averaged over a 10~Myr timescale ($\langle \mathrm{SFR} \rangle_\mathrm{10~Myr}$, see Appendix~\ref{sec:app_HeII_PopIIConfusion}), we find that the brightest Pop~III sources (with $L_\mathrm{HeII1640} \gtrsim 5 \times 10^{41} ~\si{erg.s^{-1}}$) may potentially still stand out above the Pop~II floor even in galaxies with $\langle \mathrm{SFR} \rangle_\mathrm{10~Myr} \sim 10^3 ~\si{\Msun.yr^{-1}}$.

\begin{figure}
    \centering
    \includegraphics[width=\linewidth]{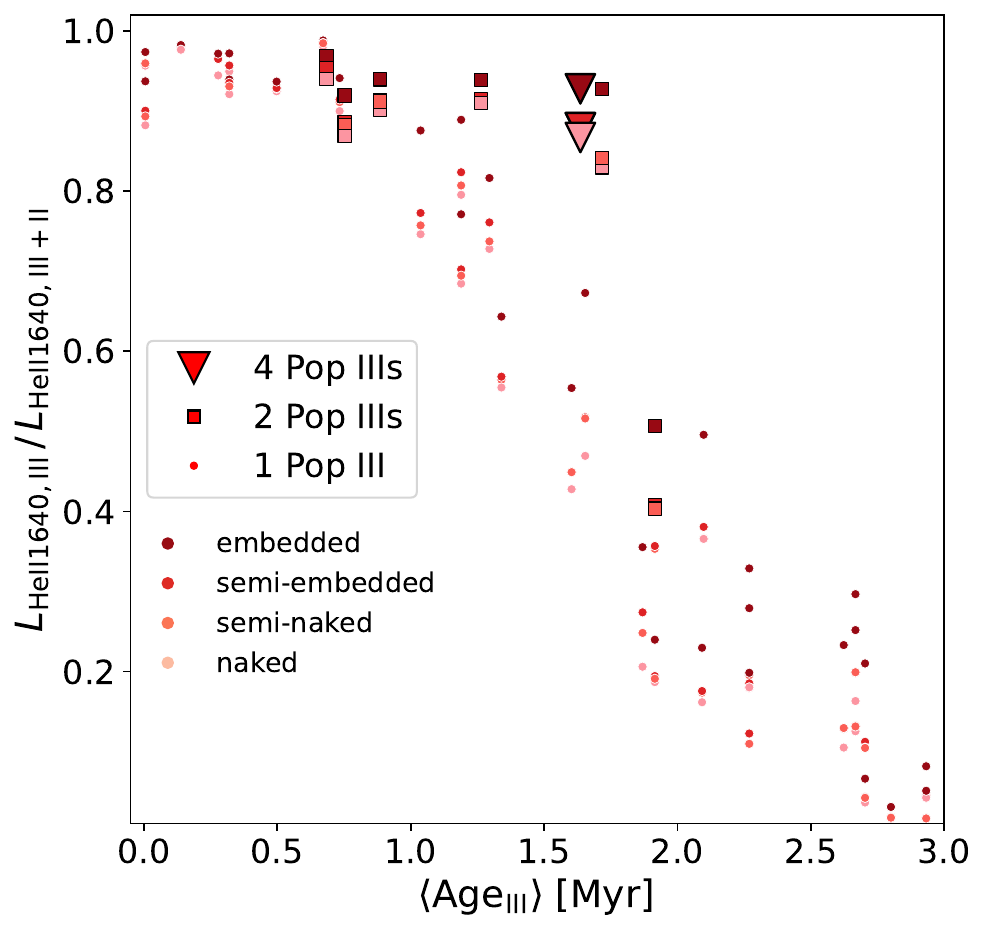}
    \caption{Fraction of the total HeII1640 line emission of Pop~III-hosting halos from Figure~\ref{fig:HeIILuminosities_haloSum} that is contributed by Pop~III stars, as a function of the mass-weighted age of the~Pop~III stellar populations in each halo. Data points are \textit{color-coded} according to the level of embedding in the assumed \texttt{Cloudy} configurations (as in Table~\ref{tab:number_density_config}). The fractional contribution of Pop~IIIs to the emission strongly decreases with age, with only a factor two scatter typically resulting from different assumed \texttt{Cloudy} configurations. All halos host a single Pop~III stellar population, with the exception of the halos shown as \textit{squares} (hosting two Pop~III populations) and the halos shown as in \textit{triangles} (hosting four~Pop~III populations).}
    \label{fig:HeIILuminosity_PopIIIFraction}
\end{figure}

These systems correspond to the halos hosting the youngest Pop~III particles in our sample, as clearly demonstrated by Figure~\ref{fig:HeIILuminosity_PopIIIFraction}. Here the fraction of the total HeII1640 line emission from the halo that is contributed by Pop~IIIs is shown as a function of the average age of the Pop~III populations. The fractional contribution of Pop~IIIs to the signal rapidly drops with age, with more than 80\% of the emission coming from Pop~III stars for the youngest systems ($\lesssim 1$~Myr), falling below 20\% for the oldest ones ($\gtrsim 2$~Myr).
This implies that not only young Pop~III populations are likely detectable in isolation (Figure~\ref{fig:HeIILuminosities_PopIIIClumps}), but they also strongly dominate the HeII signal when a substantial number of confusing Pop~II sources is present in their environment, and standard stellar populations are considered; we refer the reader to the work of Saggini et al in preparation for the discussion of more extreme assumptions including binary evolution and Wolf-Rayet stars. 

Even rare instances in which more than one alive Pop~III particle is found in the halo do not represent a significant improvement, if only relatively old, low-mass stars are left in the populations. Moreover, in most of these cases the Pop~III particles are not co-spatial, so that their emission is not likely to overlap in a random pointing (see e.g. our discussion on the prospects for spatially resolved observations in Section~\ref{sec:results_spatiallyResolvedObservations}). A notable exception is U7H4, in which four $\sim 1.5$~Myr Pop~III particles are found to lie very close to each other; this is also the only halo in the whole \texttt{dustyGadget} sample with more than two Pop~III particles within its virial radius at these redshifts.

\subsection{HI line emission and possible selection strategies}
\label{sec:results_HI}

\begin{figure*}
    \centering
    \includegraphics[width=\linewidth]{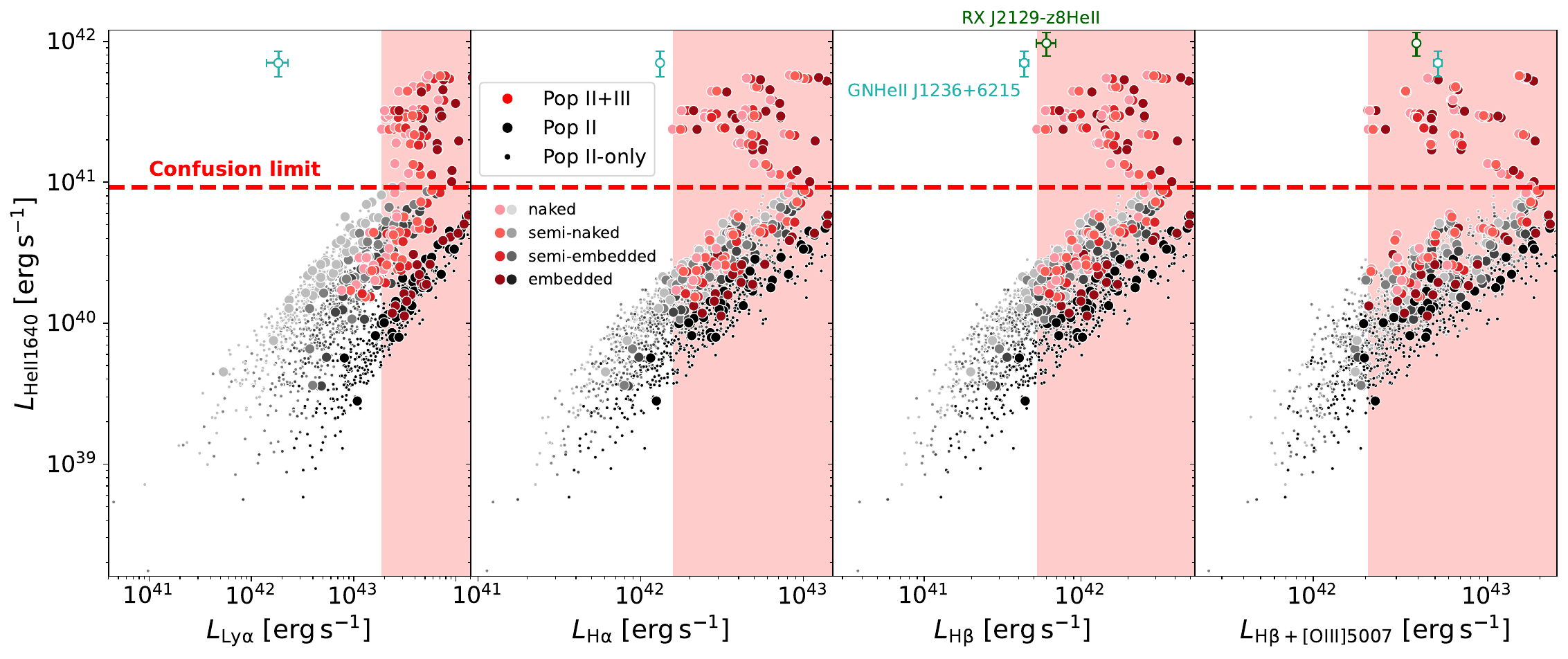}
    \caption{Integrated HeII1640 line emission ($L_\mathrm{HeII1640}$) from our Pop~III halo sample and from benchmark Pop~II-only systems (same as in Figure~\ref{fig:HeIILuminosities_haloSum}) as a function of their Ly$\alpha$ ($L_\mathrm{Ly\alpha}$, \textbf{first panel}), H$\alpha$ ($L_\mathrm{H\alpha}$, \textbf{second panel}), H$\beta$ ($L_\mathrm{H\beta}$, \textbf{third panel}) and H$\beta$ + [OIII]5007 ($L_\mathrm{H\beta + [OIII]5007}$, \textbf{last panel}) line luminosity. Data points are \textit{color-coded} according to the level of embedding in the assumed \texttt{Cloudy} configurations (as in Table~\ref{tab:number_density_config}). Pop~III systems (\textit{red-scale}) are fully degenerate with the emission from Pop~II stellar populations (\textit{grey-scale}) in terms of their HI line luminosities, making these lines alone ineffective for distinguishing the two populations. However, Pop~III sources with strong HeII emission above the confusion limit (\textit{red, dashed, horizontal line}, corresponding to the \textit{purple shaded region} in Figure~\ref{fig:HeIILuminosities_haloSum}) can still be identified. These ``unconfused'' Pop~III systems occupy the \textit{red shaded regions} of the diagram, which therefore highlights promising areas of HI luminosities for selecting potential Pop~III systems above the confusion limit. Measurements of the HeII and HI lines for the observed ``hybrid'' Pop~III candidates RXJ2129-z8HeII \citep{Wang_2024} and GNHeII-J1236+621 \citep{Mondal_2025} are also shown as a comparison.} 
    \label{fig:HeIIvsHLineLuminosity_haloSums}
\end{figure*}

Bright HI line emission has been often indicated as a potential additional tracer of Pop~III stars \citep[e.g.][]{Inoue_2011, Mas-Ribas_2016, Nakajima_Maiolino_2022}. However, our Figure~\ref{fig:HeIIvsHLineLuminosity_haloSums} (showing plots of the HeII line luminosity arising from the halos in Figure~\ref{fig:HeIILuminosities_haloSum} vs their integrated Ly$\alpha$, H$\alpha$ and H$\beta$ emission) demonstrates that, while Pop~III halos are typically found at brighter HI line luminosity than average, this emission is completely confused with standard Pop~II stellar populations; also note that HI line luminosities appear more dependent on the specific number density configuration (Table~\ref{tab:number_density_config}) than the HeII luminosities, with the regions covered by our ``naked'', ``semi-naked'', ``semi-embedded'' and ``embedded'' configurations clearly separated in the plane (these dependencies will be investigated in higher details in the paper of Graziani et al. in preparation). Therefore, strong HI lines cannot be used as conclusive evidence of the presence of Pop~IIIs in these systems, and bright HeII emission is still required for their confirmation.

On the other hand, these diagnostics may still serve as a valuable tool for selecting potential Pop~III candidates. Pop~III-forming pockets at late times are in fact expected to be extremely rare \citep{Venditti_2023}, and the stellar populations formed in these pockets will only visible for a short time (Section~\ref{sec:results_HeII_PopIIIClumps}). Moreover, searches within a large FoV around massive galaxies may be required to capture Pop~III pockets in the outskirts of the dark matter halo (see e.g. our discussion in Section~\ref{sec:results_spatiallyResolvedObservations}). Crafting selection strategies to identify promising targets for follow-up, deep searches of HeII emitters in their environment, is therefore crucial to optimize detection efficiency. 

We identify the following thresholds for Pop~III systems with HeII1640 line emission above the confusion limit (Section~\ref{sec:results_HeII_PopIIIHalos} and Figure~\ref{fig:HeIIvsHLineLuminosity_haloSums}):
\begin{equation}
    \begin{aligned}
        L_\mathrm{Ly\alpha} &\gtrsim 1.9 \times 10^{43} ~\si{erg.s^{-1}}, \\  
        L_\mathrm{H\alpha} &\gtrsim 1.5 \times 10^{42} ~\si{erg.s^{-1}}, \\
        L_\mathrm{H\beta} &\gtrsim 5.2 \times 10^{41} ~\si{erg.s^{-1}}.
    \end{aligned}
\end{equation}
Note that Ly$\alpha$ emission at these redshifts is subject to strong absorption by the neutral IGM, especially in the higher-$z$ bracket, as reionization at these redshifts is not completed. Therefore, our predictions for the $L_\mathrm{Ly\alpha}$ luminosity should be strictly regarded as upper limits.
Finally, as the H$\beta$ line is often blended with the [OIII]5007 line in observations with limited spectral resolution, the last panel of Figure~\ref{fig:HeIIvsHLineLuminosity_haloSums} shows the HeII1640 line luminosity as a function of $L_\mathrm{H\beta + [OIII]5007}$, which indicates a selection threshold for potential strong HeII emitters of
\begin{equation}
    L_\mathrm{H\beta + [OIII]5007} \gtrsim 2.0 \times 10^{42} ~\si{erg.s^{-1}}.
\end{equation}
We refer the reader to Section~\ref{sec:results_OIII} for a more extended discussion on [OIII] emission in our sample.

\subsection{Metal line emission}
\label{sec:results_OIII}

\begin{figure*}
    \centering
    \includegraphics[width=\linewidth]{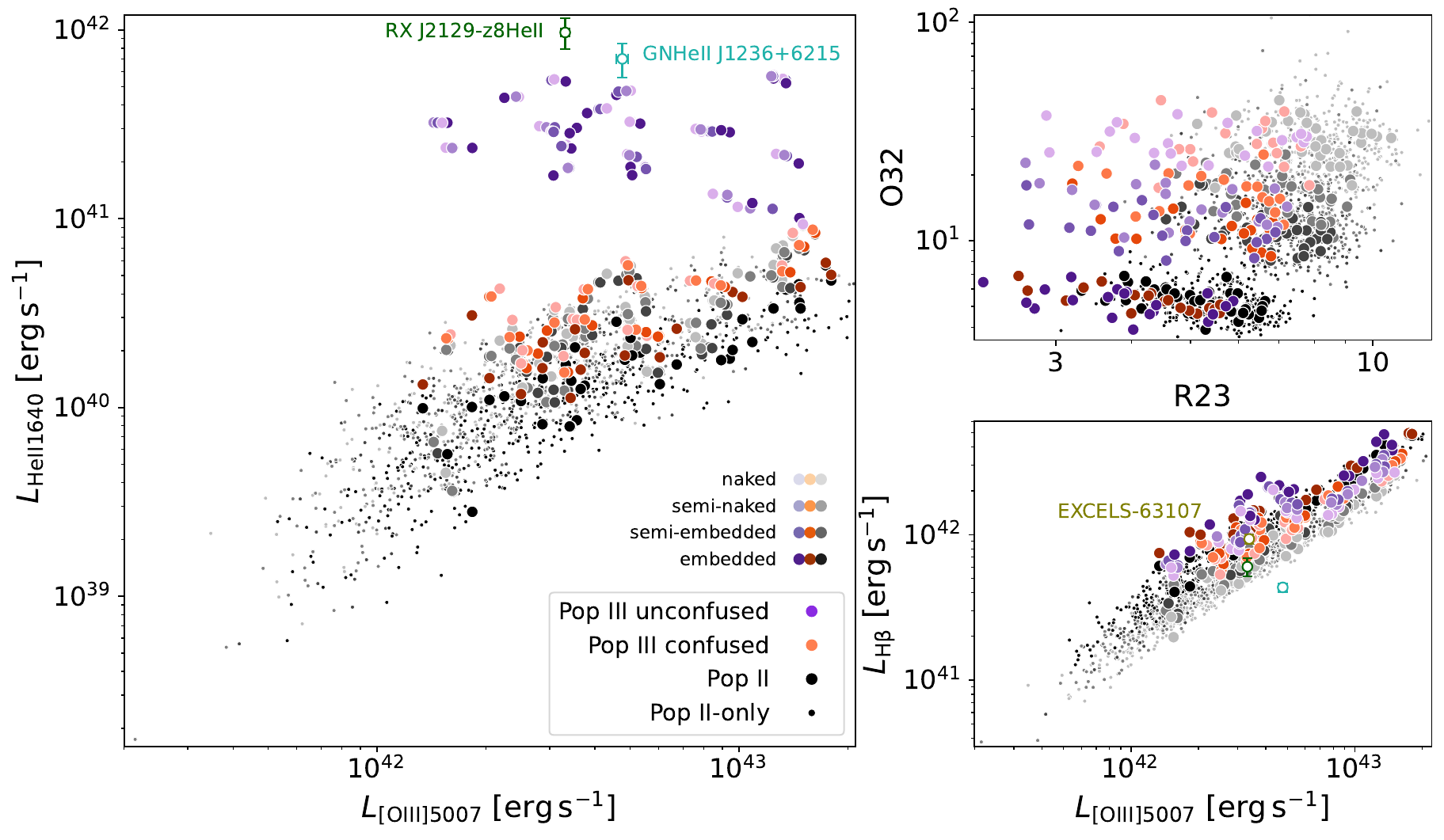}
    \caption{\textbf{Left:} Integrated HeII1640 line emission ($L_\mathrm{HeII1640}$) from our Pop~III halo sample and from benchmark Pop~II-only systems (same as in Figure~\ref{fig:HeIILuminosities_haloSum}) as a function of their [OIII]5007 line luminosity ($L_\mathrm{[OIII]5007}$), with ``unconfused'' and ``confused'' Pop~III systems (see Figure~\ref{fig:HeIILuminosities_haloSum}) highlighted in \textit{purple-} and \textit{orange-scale}, respectively. Data points are \textit{color-coded} according to the level of embedding in the assumed \texttt{Cloudy} configurations (as in Table~\ref{tab:number_density_config}). Pop~III-hosting halos are bright O line emitters, and while they are typically shifted towards the left-hand side of the O32 vs R23 optical line diagnostic plane (\textbf{top right}), this is due to their larger H$\beta$ emission, as demonstrated by their positioning in the $L_\mathrm{H\beta} - L_\mathrm{[OIII]5007}$ plane (\textbf{bottom right}). Results are compared with available constraints on the observed ``hybrid'' Pop~III candidates RXJ2129-z8HeII \citep{Wang_2024}, GNHeII-J1236+621 \citep{Mondal_2025} and EXCELS-63107 \citep{Cullen_2025}.}
    \label{fig:OIII_haloSums}
\end{figure*}

A deficit of metal lines is included in most criteria aimed at identifying metal-free star formation at high redshifts. However, residual Pop~III formation may happen within massive halos already hosting a dominant metal-enriched stellar component, as in the sample discussed in the present paper. These systems are expected to be bright metal-line emitters, due to the integrated contribution of all the underlying Pop~II populations in the halo. Figure~\ref{fig:OIII_haloSums} demonstrates this by showcasing our Pop~III halo sample against typical Pop~II-only halos in the simulations, in a plane of HeII1640 vs [OIII]5007 line luminosity: 
Pop~III-hosting systems cover a wide range of [OIII] luminosities ($\gtrsim 1$~dex), indicating no strong anti-correlation between typical tracers of the star-forming gas metallicity and the presence of pristine, star-forming clumps in the environment. On the contrary, the bulk of these systems is actually found at bright $L_\mathrm{[OIII]5007} \gtrsim 10^{42} ~\si{erg.s^{-1}}$, likely due to their typically high SFRs (see e.g. Figure~\ref{fig:MS}, and the second panel of Figure~\ref{fig:HeIILuminosity_PopIIConfusion}, as well as figure~3 of \citealt{Venditti_2023}).

The top-right panel of Figure~\ref{fig:OIII_haloSums} further showcases our Pop~II-only and Pop~III halo sample in the O32 - R23 optical diagnostic plane, respectively defined as the ratio between bright [OIII] and [OII] lines (i.e. the [OIII] lines at 5007~\AA ~and 4959~\AA ~and the [OII]3726-3728 doublet) and their cumulative flux with respect to H$\beta$, and often used as indicators of the ionization level (O32) and metallicity (R23) of galaxies. 
We see that halos hosting Pop~III stars at these redshifts appear shifted towards the left-hand side of the plane. However, this is mainly due to their bright H$\beta$ line luminosities rather than to a lower O line emission (see e.g. their positioning on the $L_\mathrm{[OIII]5007} - L_\mathrm{H\beta}$ plane in the bottom-right panel of Figure~\ref{fig:OIII_haloSums}). In fact, the star-forming gas in these halos is found at a comparable average metallicity with respect to halos of similar mass that do not host Pop~III star formation (bottom-right panel of Figure~\ref{fig:HeIILuminosity_PopIIConfusion}).

In conclusion, our results indicate that the detection of metal lines alone cannot exclude the presence of a Pop~III component in high-$z$ galaxies, and imposing a zero-metallicity constraint might actually lead us to overlook potential candidates. Note that the metal-line emission expected from Pop~III-hosting galaxies has also been explored in the work of \citet{Rusta_2025} in a lower stellar-mass regime ($M_\star \lesssim 10^{7.5} ~\Msun$), due to either recently formed Pop~II stars in the galaxy, or even rapid self-enrichment from the first Pop~III SNe in Pop~III-pure systems.
Moreover, we remark that a lack of metals is extremely difficult to establish observationally in the first place: at high redshifts, we are only able to obtain rather shallow upper limits on metallicity (see e.g. the constraints on the [OIII] line luminosities derived for the AMORE6, \citealt{Morishita_2025}, and MPG-CR3, \citealt{Cai_2025}, galaxy candidates in the bottom-right panel of Figure~\ref{fig:OIII_haloSums}), still far from the low-metallicity record established in the local Universe \citep{Ji_2025}.

\section{Discussion}
\label{sec:discussion}

\subsection{Spatially resolved observations}
\label{sec:results_spatiallyResolvedObservations}

So far we have considered integrated emission from all the stellar particles within our Pop~III-hosting halos, which corresponds to the worst-case scenario for confusion of the Pop~III signal from nearby Pop~II stars. In Section~\ref{sec:results_HeII_PopIIIHalos} we demonstrated that young, efficiently star-forming Pop~III regions could be distinguished from typical Pop~II-forming regions, even in an extreme case in which all emitting sources overlap along our LOS. In this situation, bright metal-line emission from efficiently star-forming, enriched regions would also be certainly observed (Section~\ref{sec:results_OIII}). 
Nonetheless, truly metal-free Pop~III clumps may still be accessible through spatially resolved observations. 
In Section~\ref{sec:results_HeII_PopIIIClumps} we showed that, while the identification of similar clumps in isolation would be challenging, it may still be within the reach of deep observations with JWST/NIRSpec.

\begin{figure*}
    \centering
    \includegraphics[width=0.49\linewidth]{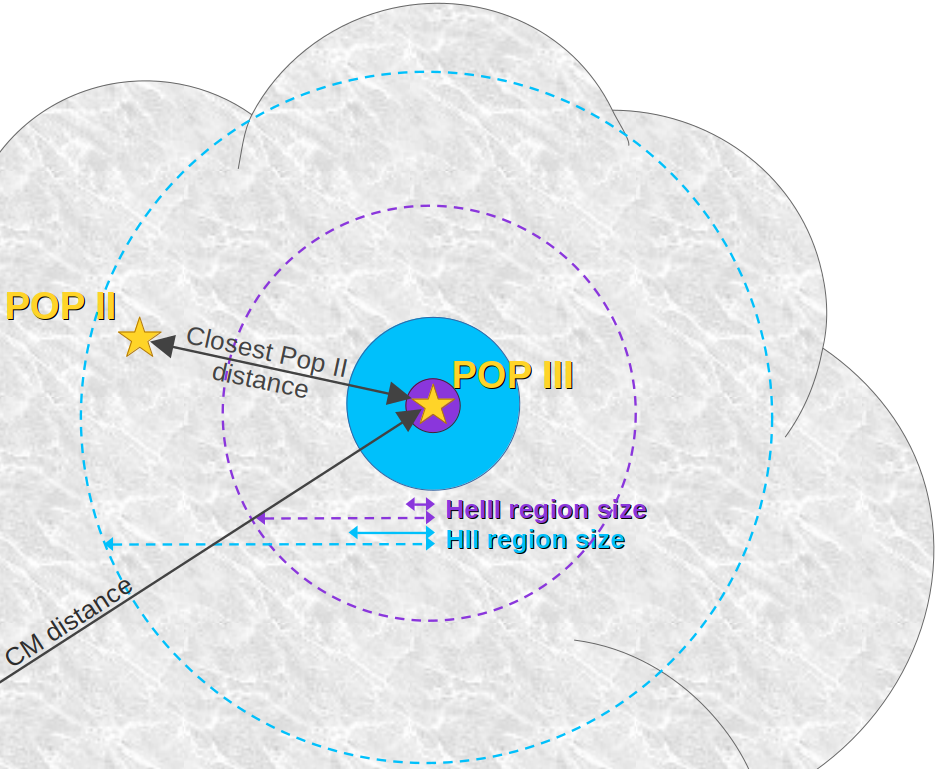}
    \includegraphics[width=0.49\linewidth]{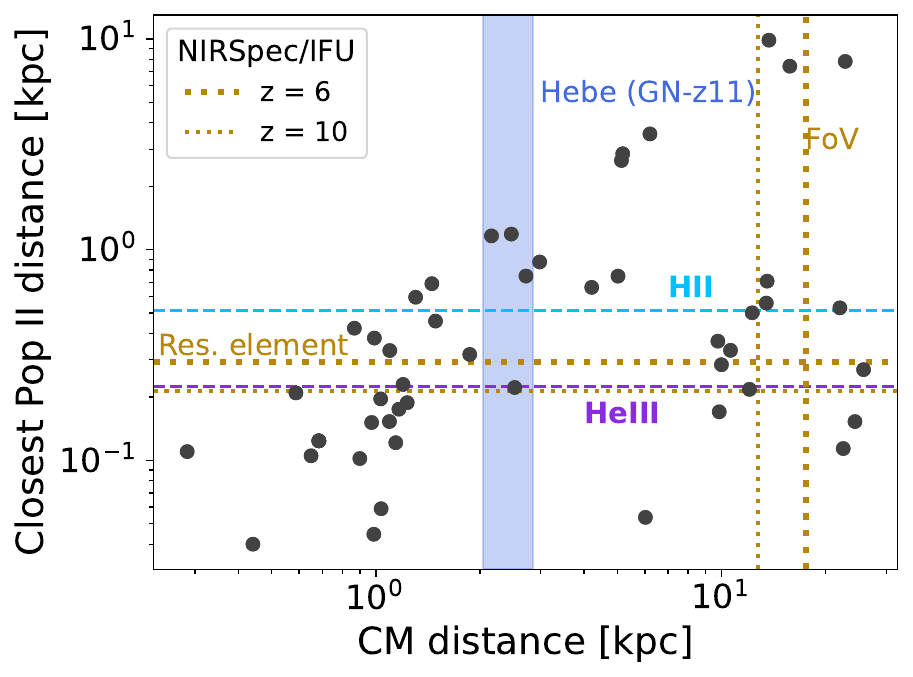}
    \caption{\textbf{Left:} scheme of a Pop~III-hosting halo, showing the distance of a Pop~III stellar particle from the center-of-mass (CM) of the halo and from its closest Pop~II particle. While all particles are physically separated in the simulation, their HII (\textit{light-blue}) and HeIII (\textit{purple}) regions may overlap, depending on the respective size of the ionized regions.
    \textbf{Right:} distance of the closest Pop~II particle as a function of the distance from the CM for all Pop~III particles in our sample. The resolution element and FoV of NIRSpec/IFU are shown as a reference at $z = 10$ and $z = 6$ (\textit{thin/thick, golden, dotted line}), as well as the inferred distance from the galaxy center of GN-z11 of the HeII clump identified by \citet{Maiolino_2024, Maiolino_2026} (\textit{blue, shaded area}). Some of the Pop~III populations in our sample are sufficiently far from nearby stars to be found in a single resolution element of NIRSpec/IFU, with no contamination from Pop~IIs. The \textit{dashed, light-blue/purple lines} indicate the maximum size of HeIII/HII regions around the Pop~III particles with our fiducial setup (see also Figure~\ref{fig:ionizedRegionSize}), showing that the emitting regions may overlap with Pop~IIs in some cases, but they would be mostly unresolved with NIRSpec/IFU.}
    \label{fig:PopIIVSCMDistances}
\end{figure*}

We explore this possibility in Figure~\ref{fig:PopIIVSCMDistances}, by comparing the distance of Pop~III particles from their closest Pop~II neighbor in our halo sample with the resolution element of NIRSpec/IFU at $z = 6$ and $z = 10$, estimated from the angular size of a single spaxel of the instrument ($\sim 0.05$~arcsec);
also note that the size of the HeII-emitting region around young Pop~III populations is typically unresolved or very close to the resolution limit (see Appendix~\ref{sec:app_HII+HeII_region_sizes}). 
This figure confirms that, while in most cases the distance of the closest Pop~II particle is lower than the resolution element, meaning that these stellar populations would be found within the same spaxel of the instrument (and the level of contamination is expected to increase even further when considering more than one spaxel in the final spectrum), a handful of the Pop~III sources considered in this study would lie sufficiently far from enriched populations to be detected in isolation, provided they are observed along a favorable LOS. In fact, Pop~III stars may be found up to $\sim 20$~kpc from the central, massive galaxy\footnote{Also see figure~10 of \citet{Venditti_2023} for a comparison of these distances with the stellar mass-weighted radius of the galaxies.}.

On the other hand, large distances from the central galaxy may cause these systems to be overlooked in observations with a limited FoV. The physical size of the FoV covered with one pointing of NIRSpec/IFU at $z = 6$ and $z = 10$ is shown as a reference in Figure~\ref{fig:PopIIVSCMDistances}, indicating that most of the systems considered in this study would be captured in a single pointing. In \citet{Venditti_2024_HeIIAnalyticalModel} we demonstrated that $\sim 90\%$ or more of the Pop~III systems predicted by our simulations may instead be missed within narrow, NIRSpec/MOS pointings centered around massive galaxies; this would have been the case, with an unlucky orientation of the instrument, for the HeII clump near the bright GN-z11 source \citep{Maiolino_2024, Maiolino_2026}, which is instead entirely within the reach of NIRSpec/IFU pointings. Even larger areas will be covered through the HARMONI spectrograph of the Extremely Large Telescope (ELT), which will) have four spaxel scales, $60 \times 30$, $20 \times 20$, $10 \times 10$ and $4 \times 4$~mas, equivalent to FoVs on the sky of $9.12 \times 6.12$, $4.08 \times 3.04$, $2.04 \times 1.52$ and $0.82 \times 0.61$~arcsec respectively\footnote{\url{https://elt.eso.org/instrument/HARMONI/}}.

Note that the level of absorption expected from intervening dust grains in the ISM (which has not been included in the present work) will also strongly depend on the position of the Pop~III sources and the morphology of the host galaxy. In \citet{Venditti_2023} we found a strong variability in the dust columns crossed by different LOS to our Pop~III sources, spanning values from $\Sigma_\mathrm{dust} \sim 10^{-3} ~\si{\Msun.kpc^{-2}}$ up to $\Sigma_\mathrm{dust} \sim 10^6 ~\si{\Msun.kpc^{-2}}$ even within a single simulated galaxy. Particularly, for Pop~III clumps found at a significant distance from the central galaxy, dust absorption may preferentially affect stellar populations lying in the most polluted regions, reducing their confusing HeII signal. However, full radiative-transfer simulations are required in order to quantify the impact of dust components in the ISM on the total emission.

\subsection{Constraining efficient Pop III-formation channels and inhomogeneous enrichment}
\label{sec:discussion_inhomogeneous_enrichment}

Our study demonstrates that massive Pop~III-hosting systems -- although intrinsically rare and likely subdominant in number -- are compelling targets. 

Theoretical studies suggest that Pop~III star formation may proceed more efficiently in atomic-cooling halos at late times (especially when immersed in the strong LW background produced by a large concentration of nearby sources, e.g. \citealt{Trinca_2026}), potentially leading to more massive starbursts and correspondingly stronger observational signatures than mini-halos \citep[e.g.][]{Greif_Bromm_2006, Greif_2008, Bromm_2009, Sugimura_2024, Jeong_2026}.
Such enhanced signals (corresponding to high values of the $\eta_\mathrm{III}$ parameter in our emission model, see Section~\ref{sec:methods_intrinsic_emission}) could remain detectable even at lower sensitivities, and be comparatively easier to identify -- even more so if occurring at intermediate redshifts; even brighter HeII luminosities may be achieved in the case of chemically homogeneous stellar evolution, for fast rotating Pop~III stars \citep{Sibony_2022, Wasserman_2026}.
Moreover, in \citet{Venditti_2023} we showed that the relative incidence of Pop~III activity may actually be higher in massive halos, despite their rarity: in fact, when pristine gas survives within overdense large-scale environments, dynamical interactions with nearby structures and satellites can promote gas compression, thereby triggering star formation within the pristine clumps \citep[e.g.][]{CorreaMagnus_2024}. 
While void regions of the cosmic web have long been the main focus of searches for late Pop~III stars, due to their slower metal enrichment \citep[e.g.][]{Rowntree_2024, Rodriguez-Medrano_2025}, these recent studies suggest that massive halos in high-density regions may actually be particularly promising targets.

As a reference, we find that in our simulations approximately $\sim 1.6 \times 10^{-5}$ Pop~III star clusters/$\si{cMpc^3}$ are hosted within large halos that also host massive star-forming galaxies ($M_\star > 10^9 ~\Msun$) at $z \approx 8.1$\footnote{The expected number and fraction of Pop~III galaxies as a function of $M_\star$ has been presented in \citet{Venditti_2023} (see e.g. their figure~3), while the match with observed populations of massive galaxies (including comparisons with the observed stellar mass function and star-forming main sequence) has been discussed in \citet{DiCesare_2023}. The expected number of candidate Pop~III systems of different mass as a function of effective survey volume, compared with JWST legacy surveys, was further explored in \citet{Venditti_2024_HeIIAnalyticalModel} (e.g. their figures~4 and~5), also taking into account potential misses due to limited FoV.}. Among these, about one fourth are younger than 1~Myr, which we expect to be bright enough to be detectable and distinguishable from the underlying Pop II floor (Sections~\ref{sec:results_HeII_PopIIIClumps} and ~\ref{sec:results_HeII_PopIIIHalos}).
We further note that, while this scenario is certainly rare, clustering analyses may provide further indications on the best environments to select when searching for similar Pop~III candidates, increasing our chances with respect to random pointings. While a full account of the environments favoring efficient Pop III star formation would require a thorough investigation of the overdensity in which our candidates lie\footnote{As a reference, in \citet{Venditti_2023} similar candidates were found within regions of the cosmic web with an average boost of more than $\sim 1000$ with respect to the average gas number density of the cubes.}, as well as the radiative background at the location of the Pop III-forming pockets in these halos, we note that all the candidates considered in the present paper are found in the vicinity of massive star-forming galaxies. In these environments, the large concentration of stellar sources is likely to boost the local radiative background, potentially to values high enough to enable the efficient Pop III formation channel pointed out by the studies of \citet{Greif_Bromm_2006, Greif_2008, Bromm_2009, Sugimura_2024, Jeong_2026}. However, this aspect certainly deserves a more extended investigation, fully accounting for radiative feedback from local stellar populations on top of the global UV background.

In addition to providing further opportunities to discover Pop~III stars beyond Cosmic Dawn, searches for recent metal-free star formation around massive galaxies will place important constraints on the topology and efficiency of metal enrichment.
Note that even models that successfully reproduce the global buildup of stellar mass at high redshifts often diverge substantially in their spatially resolved predictions, including the distribution of dust and metals in the ISM \citep[e.g.][]{Kim_2025_AGORA}. Accordingly, the very existence of pristine gas pockets in late, massive hosts remains far from settled. High-resolution zoom-in studies \citep[e.g.][]{Zier_2025, Storck_2025} suggest that efficient metal mixing may suppress Pop~III formation in the vicinity of luminous galaxies -- contrary to our predictions --, although these results remain confined to relatively limited samples. Evidence of a pervasive metal enrichment has been reported up to the highest redshifts. However, current observations may be biased towards high-metallicity regions, while overlooking pockets of very metal-poor, or even nearly primordial gas. \citet{Lewis_2025}, for example, examined potential selection bias effects favoring enriched, strong line emitters in observational studies of the mass-metallicity relation at $z \sim 3 - 6$ (also see \citealt{Kotiwale_2026} for a study of selection effects in mass-metallicity-relation estimates at $z \sim 5 - 7$). Moreover, the metallicity of high-$z$ galaxies may be overestimated when faint, metal-poor regions in their outskirts are overlooked in the presence of negative metallicity gradients \citep[e.g.][]{Li_2025}.

\subsection{Comparison with other ``hybrid'' Pop III models}
\label{sec:discussion_model_comparison}

While no study of the HeII emission of Pop~III clusters within comparatively massive halos has been performed so far, \citet{Rusta_2025} recently considered a different scenario for lower-mass ``hybrid'' Pop~III-hosting galaxies ($M_\star \lesssim 10^{7.5} ~\Msun$): in these systems, a Pop~III component may be found in moderately enriched environments, because of a surviving tail of massive stars that can still power HeII emission, even after the most massive stars formed in the current (or previous) burst have exploded as SNe (``self-polluted'' pure Pop~III phase), or in the presence of coeval Pop~II formation (truly ``hybrid'' phase). Due to the lower mass of these systems and to the different assumed IMF (i.e. a Larson-type IMF in the range $[0.8, 1000] ~\Msun$), the HeII luminosities expected from these systems are much fainter than the extreme scenario considered in the present work. For example, values up to $\approx 10^{40.1} ~\si{erg.s^{-1}}$ and $\approx 10^{39.1} ~\si{erg.s^{-1}}$ are found for the HeII1640 and HeII4686 lines respectively in the case of their ``self-polluted'' Pop~III systems with Log$U = -1$\footnote{From private communication.}.

More generally, we remark that here we considered an extreme scenario of massive Pop~III starbursts ($\sim 6 \times 10^5 ~\Msun$) with a very top-heavy IMF (Salpeter-like shape in the range $[50, 500] ~\Msun$), in order to study the detectability of a similar extreme Pop~III component even within environments dominated by Pop~II formation. While massive Pop~III starbursts within atomic-cooling halos in high-density regions of the cosmic web are predicted by simulations and consistent with recent Pop~III candidates proposed in the literature (Section~\ref{sec:methods_NEL}), a linear decrease in the HeII luminosity is expected for decreasing masses of the Pop~III clusters. The HeII emissivity also critically depends on the IMF assumption: in \citet{Venditti_2024_HeIIAnalyticalModel}, for instance, we found that the HeII emissivity at the ZAMS can vary by a factor $\sim 350$ across a broad range of Pop~III IMFs considered in the literature (see e.g. their figure~1).

\subsection{Other contaminants}
\label{sec:discussion_other_contaminants}

Finally, so far we have been focusing on the confusion arising from standard Pop~II stellar populations only, while the high temperatures required to power HeII line emission can be achieved through a number of other confusing mechanisms/sources, including X-ray binaries \citep{Schaerer_2019, Saxena_2020a, Saxena_2020b, Senchyna_2020, Cameron_2024, Lecroq_2024}, very massive stars (VMS), Wolf-Rayet (WR) stars and stripped He stars \citep{Grafener_Vink_2015, Kehrig_2018, Saxena_2020a, Shirazi_2012, Senchyna_2021, Cameron_2024, Martins_2023, Tozzi_2023, Wofford_2023, Gomez-Gonzalez_2024, Upadhyaya_2024, Berg_2025, Leitherer_2025}, AGN \citep{Saxena_2020a, Saxena_2020b, Shirazi_2012, Tozzi_2023, Liu_2024, Topping_2024}, and shocked gas \citep{Kehrig_2018, Lecroq_2024, Flury_2025}.
While a detailed study of all these contaminants has been neglected and will be deferred to future works, here we point out that strong HeII lines from AGN \citep[e.g.][]{Kollatschny_Zetzl_2013}, as well as VMS and WRs \citep[e.g.][]{Wofford_2023, Schaerer_2025, Berg_2025}, may be distinguishable from Pop~III models thanks to their larger expected line widths, although this aspect requires further investigation; particularly, \citet{Grafener_Vink_2015} showed that slow, strong winds of metal-poor ($< 0.1 ~\Zsun$) VMS may also produce narrower HeII1640 lines with FWHMs of $\sim 300 - 500 ~\si{km.s^{-1}}$. A detailed exploration of the contribution from binary stars and WRs (including estimates from the updated \texttt{X-BPASS} model, \citealt{Bray_2025_X-BPASS}) will be presented in Saggini et al. in preparation, although see Figure~\ref{fig:HeIILuminosity_PopIIConfusion} and Appendix~\ref{sec:app_HeII_PopIIConfusion} for a comparison with other estimates of the WR contribution to the HeII1640 line emission from \citet{Cassata_2013} and \citet{Visbal_2015}.

\section{Conclusions}
\label{sec:conclusions}

We have investigated the nebular line-emission signatures of a subdominant Pop~III stellar component in the environment of massive ($M_\star \gtrsim 10^9 ~\Msun$), Pop II-dominated galaxies at $z \simeq 6.5 - 9$. We combined a halo sample from the \texttt{dustyGadget} cosmological simulation suite with a spatially resolved (``particle-by-particle'') photoionization post-processing framework, coupling the SPS models \texttt{BPASS} and \texttt{Yggdrasil} with the photoionization code \texttt{Cloudy}.

Our main result is that young and sufficiently massive Pop III-forming clumps can still produce a distinctive observational signature, even within globally enriched environments. Particularly, strong HeII1640 line emission is confirmed as the most robust tracer of ongoing Pop~III activity: recombinations around Pop~III clusters with ages $\lesssim 1$~Myr, a top-heavy IMF and masses of order $\sim 6 \times 10^5 ~\Msun$ can reach $L_\mathrm{HeII1640} \gtrsim 10^{41} ~\si{erg.s^{-1}}$, making them detectable with $\sim 10$~h of observations with JWST/NIRSpec up to $z = 10$. On the other hand, such luminosities cannot be reproduced by standard Pop~II populations alone.

At the same time, the broader environment of these massive ``hybrid'' hosts is expected to be strongly contaminated by Pop~II emission. Bright metal lines (e.g. $L_\mathrm{[OIII]5007} \sim 10^{42} - 2 \times 10^{43} ~\si{erg.s^{-1}}$) are naturally powered by the dominant enriched stellar component, implying that the detection of metal lines does not by itself rule out the presence of Pop~III stars in these systems. Likewise, strong Ly$\alpha$, H$\alpha$, and H$\beta$ emission (i.e. $L_\mathrm{Ly\alpha} \gtrsim 1.9 \times 10^{43} ~\si{erg.s^{-1}}$, $L_\mathrm{H\alpha} \gtrsim 1.5 \times 10^{42} ~\si{erg.s^{-1}}$, and $L_\mathrm{H\beta} \gtrsim 5.2 \times 10^{41} ~\si{erg.s^{-1}}$) proved to be a useful candidate-selection tool, but not sufficient to uniquely identify Pop~III activity, due to confusion with Pop~II emission.

Overall, our analysis suggests that searching for spatially segregated, short-lived HeII-bright pockets in the outskirts of massive halos is a promising path to uncover residual Pop~III formation at the end of reionization.

\begin{acknowledgments}
We thank Elka Rusta and Stefania Salvadori for providing predictions of the HeII line luminosities from the \texttt{NEFERTITI} model.
AV acknowledges funding from the Cosmic Frontier Center and the University of Texas at Austin’s College of Natural Sciences.
AV, LG and RS acknowledge support from the PRIN 2022 MUR project 2022CB3PJ3 - First Light And Galaxy aSsembly (FLAGS) funded by the European Union – Next Generation EU. 
JBM was supported by NSF Grants AST-2307354 and AST-2408637, and by the NSF-Simons AI Institute for Cosmic Origins.
This research was also supported in part by grant NSF PHY-2309135 to the Kavli Institute for Theoretical Physics (KITP). 
RV acknowledges support from PRIN MUR "2022935STW" funded by European Union-Next Generation EU, Missione 4 Componente 2 CUP C53D23000950006 and from Bando Ricerca Fondamentale INAF 2023, Theory Grant "Theoretical models for Black Holes Archaeology". CDC acknowledges support from the European Union (ERC, AGENTS, 101076224).
\end{acknowledgments}

\software{
\texttt{dustyGadget} \citep{Graziani_2020},
\texttt{\href{https://zenodo.org/records/6338460}{BPASSv2.2.1}} \citep{Eldridge_2017, Stanway_Elridge_2018},
\texttt{\href{https://www.astro.uu.se/~ez/yggdrasil/yggdrasil.html}{Yggdrasil}} \citep{Zackrisson_2011},
\texttt{\href{https://gitlab.nublado.org/cloudy/cloudy}{Cloudy22.01}} \citep{Ferland_2017},
\texttt{\href{https://numpy.org}{numpy}} \citep{VanDerWalt_2011_numpy, Harris_2020_numpy},
\texttt{\href{https://matplotlib.org}{matplotlib}} \citep{Hunter_2007_matplotlib}, 
\texttt{\href{https://scipy.org}{scipy}} \citep[\href{https://mail.python.org/pipermail/python-list/2001-August/106419.html}{Jones et al. 2001};][]{Virtanen_2020_scipy}.
}

\appendix

\section{Pop~III halo sample physical properties}
\label{sec:app_PopIII_halo_properties}

Table~\ref{tab:PopIII_halo_sample} summarizes the physical properties of the Pop~III halos considered in this work.

\begin{table*}
    \centering
    \begin{tabular}{cccccccc}
        \hline
         Halo ID & 
         $\mathrm{Log} (M_\mathrm{vir} / \Msun)$ & 
         $\mathrm{Log} (M_\star / \Msun)$ & $\langle \mathrm{SFR} \rangle_\mathrm{100~Myr}$ & $\langle \mathrm{SFR} \rangle_\mathrm{10~Myr}$ & $\langle \mathrm{Age_\star} \rangle_\mathrm{10~Myr}$ & 
         $\langle Z_\star \rangle_\mathrm{10~Myr}$ \\
         & & ($M_\mathrm{III}/M_\star$) & [$\si{\Msun.yr^{-1}}$] & [$\si{\Msun.yr^{-1}}$] & [Myr] & [$\Zsun$] \\
        \hline
        \multicolumn{7}{c}{$z = 9.0$} \\ 
        \hline
        U10H0 & 11.22 & 8.92 (-2.60) & 6.8 & 11.2 & 4.97 & 0.034 \\
        \hline 
        \multicolumn{7}{c}{$z = 8.1$} \\ 
        \hline
        U6H5 & 11.19 & 8.97 (-2.65) & 5.7 & 7.4 & 4.49 & 0.059 \\
        U6H8 & 11.17 & 8.69 (-2.37) & 4.0 & 8.7 & 5.32 & 0.029 \\
        U7H4 & 11.37 & 9.18 (-2.26) & 11.4 & 18.5 & 3.78 & 0.049 \\
        U7H8 & 11.15 & 9.05 (-2.73) & 8.4 & 11.9 & 4.82 & 0.062 \\
        U8H1 & 11.51 & 9.33 (-2.93) & 16.3 & 26.9 & 5.14 & 0.046 \\
        U8H2 & 11.46 & 9.37 (-3.06) & 15.9 & 21.0 & 4.82 & 0.055 \\
        U8H5 & 11.27 & 9.02 (-2.61) & 7.3 & 10.8 & 4.29 & 0.050 \\
        U10H2 & 11.26 & 9.01 (-2.70) & 7.1 & 11.5 & 4.18 & 0.040 \\
        U10H5 & 11.06 & 8.99 (-2.68) & 6.6 & 7.2 & 4.23 & 0.091 \\
        U12H1 & 11.45 & 9.35 (-3.04) & 16.0 & 22.1 & 4.64 & 0.059 \\
        U12H3 & 11.27 & 9.40 (-3.08) & 15.1 & 18.1 & 5.34 & 0.087 \\
        \hline 
        \multicolumn{7}{c}{$z = 7.3$} \\ 
        \hline
        U6H0 & 11.77 & 10.00 (-3.59) & 59.9 & 73.2 & 5.03 & 0.131 \\
        U6H2 & 11.47 & 9.44 (-3.13) & 14.3 & 17.5 & 4.92 & 0.058 \\
        U6H10 & 11.33 & 9.18 (-2.87) & 10.0 & 14.8 & 4.66 & 0.052 \\
        U7H2 & 11.64 & 9.62 (-2.91) & 27.4 & 51.9 & 4.57 & 0.063 \\
        U7H3 & 11.62 & 9.44 (-2.83) & 16.0 & 16.7 & 4.65 & 0.049 \\
        U8H0 & 11.92 & 9.99 (-3.68) & 62.6 & 86.6 & 5.03 & 0.094 \\
        U8H1 & 11.78 & 9.83 (-3.52) & 41.0 & 63.1 & 4.82 & 0.079 \\
        U8H4 & 11.48 & 9.49 (-3.18) & 17.1 & 26.8 & 4.81 & 0.080 \\
        U10H2 & 11.47 & 9.41 (-2.80) & 15.7 & 21.4 & 4.35 & 0.061 \\
        U10H3 & 11.43 & 9.33 (-3.02) & 11.8 & 12.7 & 5.01 & 0.076 \\
        U12H2 & 11.64 & 9.50 (-3.19) & 19.1 & 29.5 & 5.29 & 0.058 \\
        U13H0 & 11.54 & 9.35 (-2.69) & 14.1 & 18.6 & 5.01 & 0.055 \\
        U13H2 & 11.43 & 9.38 (-3.07) & 14.2 & 16.2 & 4.99 & 0.068 \\
        \hline 
        \multicolumn{7}{c}{$z = 6.5$} \\ 
        \hline
        U6H5 & 11.57 & 9.81 (-3.14) & 33.7 & 39.9 & 4.81 & 0.133 \\
        U6H6 & 11.58 & 9.74 (-3.43) & 25.3 & 36.8 & 4.80 & 0.093 \\
        U6H15 & 11.51 & 9.65 (-3.33) & 23.0 & 30.0 & 5.09 & 0.112 \\
        U6H17 & 11.47 & 9.59 (-3.28) & 18.9 & 22.9 & 4.45 & 0.110 \\
        U7H3 & 11.81 & 9.85 (-3.19) & 39.8 & 64.4 & 5.02 & 0.066 \\
        U7H4 & 11.78 & 9.87 (-3.46) & 42.8 & 73.2 & 5.20 & 0.084 \\
        U7H20 & 11.43 & 9.40 (-3.09) & 13.2 & 15.6 & 4.99 & 0.084 \\
        U10H1 & 11.76 & 9.89 (-3.58) & 40.4 & 49.4 & 5.11 & 0.117 \\
        U10H4 & 11.63 & 9.63 (-3.32) & 17.2 & 28.1 & 6.02 & 0.051 \\
        U12H2 & 11.79 & 9.90 (-3.59) & 43.0 & 76.8 & 4.95 & 0.088 \\
        U12H3 & 11.70 & 9.90 (-3.59) & 43.2 & 44.7 & 4.60 & 0.128 \\
        U12H9 & 11.58 & 9.69 (-3.38) & 24.7 & 35.3 & 4.58 & 0.096 \\
        U12H16 & 11.48 & 9.56 (-3.15) & 16.8 & 21.9 & 5.36 & 0.094 \\
        U12H36 & 11.34 & 9.45 (-3.14) & 9.4 & 14.3 & 5.46 & 0.082 \\
        \hline
    \end{tabular}
    \caption{List of the Pop~III halos considered in this study and their physical properties: virial mass of the halo ($M_\mathrm{vir}$), total stellar mass ($M_\star$), Pop~III stellar mass fraction ($M_\mathrm{III} / M_\star$), total SFR averaged over 100 Myr ($\langle \mathrm{SFR} \rangle_\mathrm{100~Myr}$) and 10 Myr ($\langle \mathrm{SFR} \rangle_\mathrm{10~Myr}$), and mass-weighted age ($\langle \mathrm{Age_\star} \rangle_\mathrm{10~Myr}$) and metallicity ($\langle Z_\star \rangle_\mathrm{10~Myr}$) of all stellar populations younger than 10 Myr.}
    \label{tab:PopIII_halo_sample}
\end{table*}

\section{Confusion from Pop~II stars and dependence on galaxy properties}
\label{sec:app_HeII_PopIIConfusion}

\begin{figure*}
    \centering
    \includegraphics[width=\linewidth]{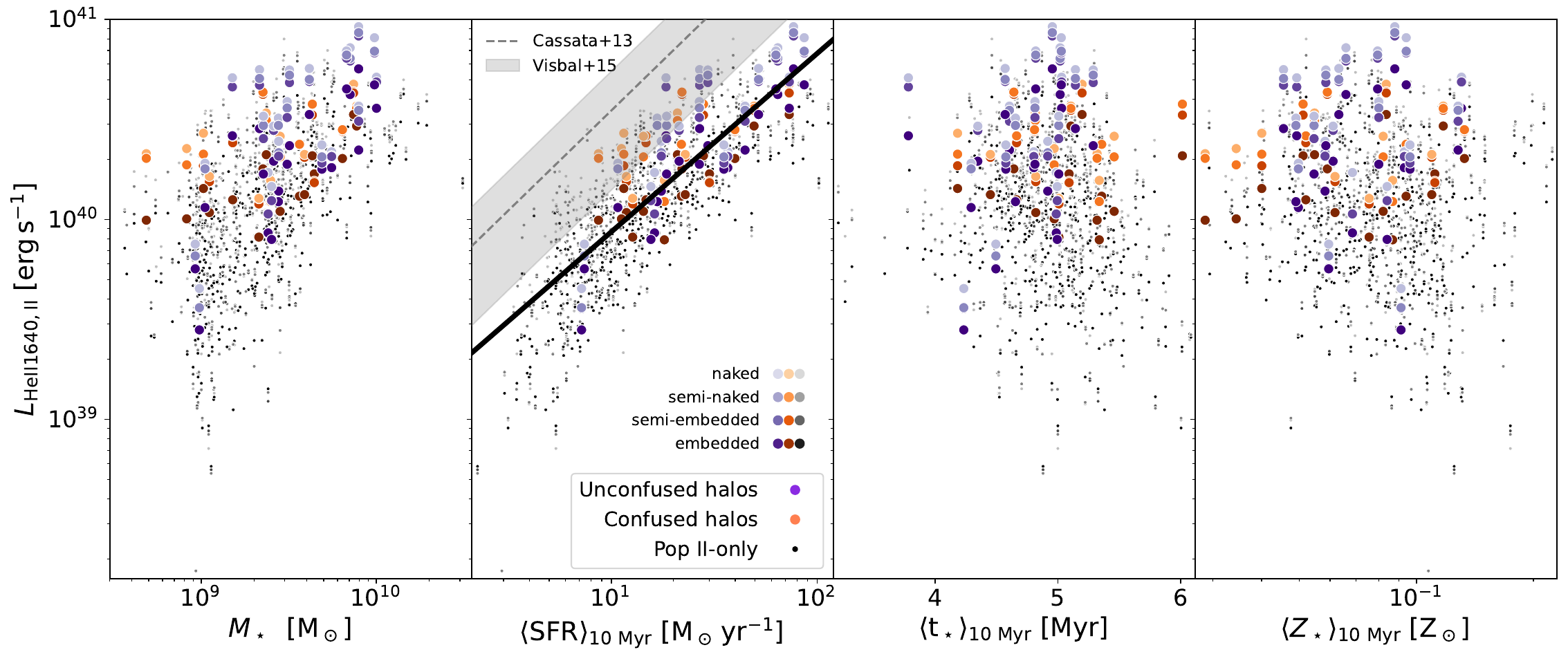}
    \caption{Integrated contribution to the HeII1640 line emission coming from Pop~II stellar populations alone ($L_\mathrm{HeII1640,II}$) within our Pop~III halo sample and benchmark Pop~II-only systems (same as in Figure~\ref{fig:HeIILuminosities_haloSum}) as a function of total stellar mass ($M_\star$, \textbf{first panel}) and SFR averaged over 10~Myr (\textbf{second panel}), and as a function of mass-averaged stellar population properties (stellar age $t_\star$ and metallicity $Z_\star$, \textbf{third panel} and \textbf{fourth panel} respectively) for the emitting stellar populations ($< 10$~Myr of age). Data points are \textit{color-coded} according to the level of embedding in the assumed \texttt{Cloudy} configurations (as in Table~\ref{tab:number_density_config}). The contamination level on the HeII signal from standard Pop~II population mainly depends on the amount of young stellar mass in the halos, with no apparent dependence on the properties of the emitting populations within the considered range. The \textit{black solid line} in the \textbf{second panel} shows a linear fit of the relation between $L_\mathrm{HeIII1640,II}$ for our fiducial \texttt{Cloudy} models (Section~\ref{sec:methods_NEL}) and the SFR of the halo averaged over 10~Myr (see text for details), while the \textit{grey dashed line} and the \textit{grey shaded area} respectively show a linear fit from the empirical $L_\mathrm{HeII1640,WR}$ vs SFR relation found by \citet{Cassata_2013} for the HeII1640 line emission from WR stars in galaxies at $2 < z < 4.6$ (with SFR inferred from SED fitting), and the analogous $L_\mathrm{HeII1640,WR}$ vs SFR relation derived by \citet{Visbal_2015} by adopting the model of \citet{Schaerer_2003} (assuming a Salpeter IMF in the range $[1, \, 100] ~\Msun$ and constant SFR, at $10^{-3} ~\Zsun$). Pop~III-hosting systems are highlighted in \textit{color}, with ``unconfused'' and ``confused'' Pop~III systems (see Figure~\ref{fig:HeIILuminosities_haloSum}) highlighted in \textit{purple-} and \textit{orange-scale}, respectively.}
    \label{fig:HeIILuminosity_PopIIConfusion}
\end{figure*}

The impact of our considered \texttt{Cloudy} configurations in determining the level of confusing emission from Pop~II stars appears fairly limited, only yielding a factor two spread in the fractional contribution to the emission in Figure~\ref{fig:HeIILuminosity_PopIIIFraction}. 
The total HeII emission arising from Pop~II stars mainly depends on the total young stellar mass ($< 10$~Myr), as demonstrated by the strongly increasing trend of the HeII1640 line luminosity with the SFR averaged over 10 Myr in Figure~\ref{fig:HeIILuminosity_PopIIConfusion}, while the dependence on average stellar population properties such as age and metallicity is much weaker in comparison\footnote{Note that while the average age and metallicity of the stellar populations in each of the Pop~III-hosting halos are indicated for reference in the last two panels of Figure~\ref{fig:HeIILuminosity_PopIIConfusion}, in the \texttt{Cloudy} runs we take into account the different ages of each individual stellar population, as predicted by the simulation at the redshift of interest, rather than assigning an average effective age to the whole stellar component (as e.g. in \citealt{Charlot_Longhetti_2001} and follow-up works).}. 
On the other hand, Pop~III stars are predominantly found in highly star-forming systems\footnote{Also see our figure~3 in \citet{Venditti_2023}, in which the main sequence of star formation for Pop~III halos vs Pop~III-only halos is shown for the entire halo population in the dustyGadget sample.  Pop~III halos are mostly found at the high end of the sequence at a given stellar mass, suggesting that the same conditions that favor star formation in the pristine clumps within these halos may also favor enhanced metal-enriched star formation in the enriched regions.}, with fairly large amounts of young stellar populations, indicating a potential trade-off between the expected level of confusion from Pop~II stars and the probability of finding Pop~III-forming regions in the environment, when selecting massive galaxy candidates for follow-up.

At lower redshifts, the contamination level may be higher as galaxies grow more massive. On the other hand, only fairly young stellar populations contribute to the HeII emission. As our results are found to be largely independent from the average properties of the emitting stellar populations, a first-order estimate of the expected level of confusion from enriched populations in massive, star-forming halos may be obtained by extrapolating our $L_\mathrm{HeII1640}$ vs SFR relation at $z \sim 6 - 9$. To this aim, the third panel of Figure~\ref{fig:HeIILuminosity_PopIIConfusion} shows a linear fit of this relation for our fiducial \texttt{Cloudy} models (Section~\ref{sec:methods_NEL}), considering the SFR averaged over 10~Myr: 
\begin{equation}
    \mathrm{Log} \left(\frac{L_\mathrm{HeII1640}}{\si{erg.s^{-1}}}\right) = 
    (39.062 \pm 0.020) + (0.879 \pm 0.017) \times
    \mathrm{Log} \left(\frac{\mathrm{\langle SFR \rangle}_\mathrm{10~\si{Myr}}}{\si{\Msun.yr^{-1}}}\right).
\end{equation}
If we assume that at best one or two Pop~III-forming clumps may be found in low-redshift halos, we expect that clumps with a HeII luminosity in the order of $L_\mathrm{HeII1640} \sim 5 \times 10^{41} ~\si{erg.s^{-1}}$ (as e.g. the youngest cluster in Figure~\ref{fig:HeIILuminosities_PopIIIClumps}) should be distinguishable from the underlying Pop~II emission up to a large SFR of $\sim 10^3 ~\si{\Msun.yr^{-1}}$. However we caution the reader against the extrapolation of our relation beyond the average stellar-metallicity range considered in this study ($\sim 10^{-2} - 3 \times 10^{-1} ~\Zsun$). We also remind that the gas properties in the vicinity of the stellar populations (including its number density, dust-to-metal ratio and metallicity boost with respect to the stellar metallicity) have been calibrated against observations of typical star-forming galaxies in the $z \sim 6 - 9$ redshift range (Graziani et al. in preparation), therefore, any extrapolation beyond this range should be taken \textit{cum grano salis}. For example, in the second panel of Figure~\ref{fig:HeIILuminosity_PopIIConfusion} we show the empirical $L_\mathrm{HeII1640,WR}$ vs SFR relation found by \citet{Cassata_2013} for the HeII1640 line emission from WR stars in galaxies at $2 < z < 4.6$ (with SFR inferred from SED fitting), and the same relation derived by \citet{Visbal_2015} by adopting the model of \citet{Schaerer_2003} at $10^{-3} ~\Zsun$ (assuming a Salpeter IMF in the range $[1, \, 100] ~\Msun$ and constant SFR), demonstrating that lower metallicities, as well as different galaxy properties at lower redshifts, may raise the expected contamination level substantially. Different modeling choices for the WR and binary components will be explored in Saggini et al. in preparation.

\section{Typical HII and HeII region sizes of Pop~III star-forming regions}
\label{sec:app_HII+HeII_region_sizes}

\begin{figure*}
    \centering
    \includegraphics[width=0.55\linewidth]{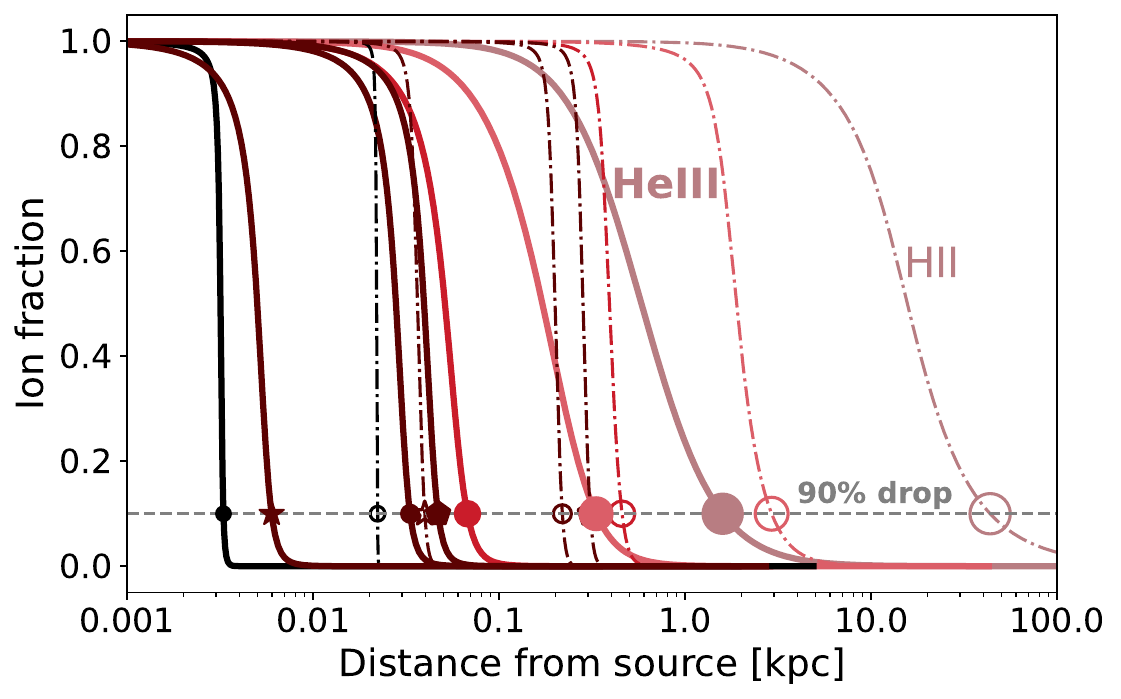}
    \includegraphics[width=0.44\linewidth]{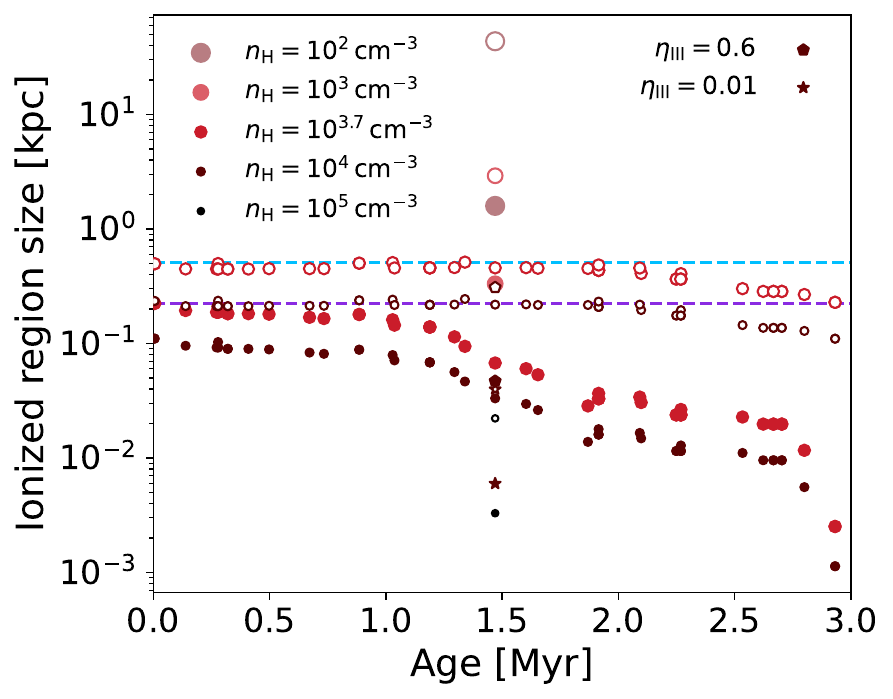}
    \caption{\textbf{Left:} HeIII (\textit{thick, solid lines}) and HII (\textit{thin, dash-dotted lines}) fraction around a Pop~III particle of $\approx 1.5$~Myr as a function of the distance from the source, adopting the same exploration of the physical setup of Figure~\ref{fig:HeIILuminosities_PopIIIClumps}, and two additional cases with $\eta_\mathrm{III} = 0.6$ and 0.01, assuming $n_\mathrm{H} = 10^4$ (\textit{dark-red pentagon/star}). A smoother, slower decay of the ion fractions with distance is found for decreasing number densities ($n_\mathrm{H} = 10^5 \rightarrow 10^2 ~\si{cm^{-3}}$) and increasing Pop~III masses ($\eta_\mathrm{III} = 0.01 \rightarrow 0.6$, corresponding to $M_\mathrm{III} = 2 \times 10^4 \rightarrow 1.2 \times 10^6 ~\Msun$).
    \textbf{Right:} size of the HeIII (\textit{filled points}) and HII (\textit{empty points}) regions (estimated at a 90\% drop of the ionized fraction, see left panel) for all Pop~III particles in our sample as a function of their age, as in Figure~\ref{fig:HeIILuminosities_PopIIIClumps}. The youngest Pop~III populations, i.e. the brightest HeII emitters (most likely to be observable, Figure~\ref{fig:HeIILuminosities_PopIIIClumps}), also have the largest HeIII/HII regions (indicated by \textit{light-blue/purple horizontal dashed lines} respectively), and are therefore more likely to overlap with nearby Pop~IIs.}
    \label{fig:ionizedRegionSize}
\end{figure*}

As Pop~III HeII emitters are not point-like, the size of the HeII-emitting region needs to be taken into account when discussing spatially resolved observations, as in our Section~\ref{sec:results_spatiallyResolvedObservations}. 
To this aim, the left panel of Figure~\ref{fig:ionizedRegionSize} shows the HeIII fraction in the region surrounding a single $\sim 1.5$~Myr Pop~III stellar particle as a function of the distance from the source, with the same physical setup exploration of Figure~\ref{fig:HeIILuminosities_PopIIIClumps}, as well as two additional cases with $\eta_\mathrm{III} = 0.6$ and 0.01, assuming $n_\mathrm{H} = 10^4$: the ionized fraction quickly drops within $\sim 10$~pc in the case of dense, optically thick clouds (e.g. $n_\mathrm{H} = 10^5 ~\si{cm.^{-3}}$) or less massive, weakly ionizing sources (e.g. with $\eta_\mathrm{III} = 0.01$, i.e. Pop~III masses of $\sim 2 \times 10^4 ~\Msun$), while much more extended HeIII regions (even up to a $\sim$~kpc scale) are found around massive Pop~III populations (e.g. $\eta_\mathrm{III} = 0.3 - 0.6$, i.e. $M_\mathrm{III} \sim 0.6 - 1.2 \times 10^5 ~\Msun$) embedded in low-density gas (e.g. $n_\mathrm{H} = 10^2 ~\si{cm.^{-3}}$). The size of the HeII-emitting region (estimated as the region where the HeIII fraction drops below 10\% of its initial value) for all the other Pop~III particles of Figure~\ref{fig:HeIILuminosities_PopIIIClumps} in our fiducial setup ($\eta_\mathrm{III} = 0.3$, $n_\mathrm{H} = 10^{3.7} - 10^4 ~\si{cm^{-3}}$) is also shown in the right panel of Figure~\ref{fig:ionizedRegionSize} as a function of their age\footnote{Note that smaller HII/HeIII regions are expected around Pop~II particles, e.g. with typical HII regions of sizes up to $\sim 50$~pc.}. The youngest populations ($\lesssim 1$~Myr) may yield emitting regions up to a few hundreds of pc, which are large enough to blend with neighboring Pop~IIs for some of the Pop~III systems, but still small enough to be unresolved with JWST/NIRSpec at $z \approx 6 - 10$. However, these predictions appear much more dependent on the physical setup than the corresponding predictions for the HeII line luminosity (Figure~\ref{fig:HeIILuminosities_PopIIIClumps}).
Also note that, in comparison, the HII regions are much larger even around older populations, as lower-energy photons are sufficient to keep H ionized up to higher distances.

\clearpage

\bibliography{main}{}
\bibliographystyle{aasjournalv7}

\end{document}